\documentclass[longauth]{aaEC}

\usepackage{graphicx}
\usepackage{natbib}
\usepackage{scalerel}
\usepackage{lastpage}
\usepackage[table]{xcolor}

\usepackage{txfonts}
\usepackage{hyperref}
\hypersetup{
    colorlinks=true,
    linkcolor=blue,
    filecolor=magenta,      
    urlcolor=blue,
    citecolor=blue
}

\makeatletter
\renewcommand*\aa@pageof{, page \thepage{} of \pageref*{LastPage}}
\makeatother

\usepackage[switch, modulo]{lineno}

\usepackage{euclid}

\newcommand{\oiii}[1]{[\ion{O}{iii}]#1}
\newcommand{\oii}[1]{[\ion{O}{ii}]#1}

\newcommand{\nii}[1]{[\ion{N}{ii}]#1}

\newcommand{\siii}[1]{[\ion{S}{iii}]#1}

\newcommand{\esc}{\ensuremath{\mathrm{erg\,s}^{-1}\,\mathrm{cm}^{-2}}}

\newcommand{\Pa}{\ensuremath{\mathrm{Pa}\,\alpha}}

\newcommand{\ecgs}
{\ensuremath{\mathrm{erg\,s}^{-1}\,\mathrm{cm}^{-2}}}

\begin{document}
\title{Euclid Quick Data Release (Q1)}
\subtitle{Characteristics and limitations of the spectroscopic measurements\thanks{This paper is dedicated to the memory of Bianca Garilli, an unwavering supporter of this survey.}}

\newcommand{\orcid}[1]{} 
\author{Euclid Collaboration: V.~Le~Brun\orcid{0000-0002-5027-1939}\thanks{\email{Vincent.Lebrun@lam.fr}}\inst{\ref{aff1}}
\and M.~Bethermin\orcid{0000-0002-3915-2015}\inst{\ref{aff2}}
\and M.~Moresco\orcid{0000-0002-7616-7136}\inst{\ref{aff3},\ref{aff4}}
\and D.~Vibert\orcid{0009-0008-0607-631X}\inst{\ref{aff1}}
\and D.~Vergani\orcid{0000-0003-0898-2216}\inst{\ref{aff4}}
\and C.~Surace\orcid{0000-0003-2592-0113}\inst{\ref{aff1}}
\and G.~Zamorani\orcid{0000-0002-2318-301X}\inst{\ref{aff4}}
\and A.~Allaoui\inst{\ref{aff1}}
\and T.~Bedrine\inst{\ref{aff1}}
\and P.-Y.~Chabaud\inst{\ref{aff1}}
\and G.~Daste\inst{\ref{aff1}}
\and F.~Dufresne\inst{\ref{aff1}}
\and M.~Gray\inst{\ref{aff1}}
\and E.~Rossetti\orcid{0000-0003-0238-4047}\inst{\ref{aff5}}
\and Y.~Copin\orcid{0000-0002-5317-7518}\inst{\ref{aff6}}
\and S.~Conseil\orcid{0000-0002-3657-4191}\inst{\ref{aff6}}
\and E.~Maiorano\orcid{0000-0003-2593-4355}\inst{\ref{aff4}}
\and Z.~Mao\orcid{0000-0003-4016-845X}\inst{\ref{aff4}}
\and E.~Palazzi\orcid{0000-0002-8691-7666}\inst{\ref{aff4}}
\and L.~Pozzetti\orcid{0000-0001-7085-0412}\inst{\ref{aff4}}
\and S.~Quai\orcid{0000-0002-0449-8163}\inst{\ref{aff3},\ref{aff4}}
\and C.~Scarlata\orcid{0000-0002-9136-8876}\inst{\ref{aff7}}
\and M.~Talia\orcid{0000-0003-4352-2063}\inst{\ref{aff3},\ref{aff4}}
\and H.~M.~Courtois\orcid{0000-0003-0509-1776}\inst{\ref{aff8}}
\and L.~Guzzo\orcid{0000-0001-8264-5192}\inst{\ref{aff9},\ref{aff10},\ref{aff11}}
\and B.~Kubik\orcid{0009-0006-5823-4880}\inst{\ref{aff6}}
\and A.~M.~C.~Le~Brun\orcid{0000-0002-0936-4594}\inst{\ref{aff12}}
\and J.~A.~Peacock\orcid{0000-0002-1168-8299}\inst{\ref{aff13}}
\and D.~Scott\orcid{0000-0002-6878-9840}\inst{\ref{aff14}}
\and D.~Bagot\inst{\ref{aff15}}
\and A.~Basset\inst{\ref{aff15}}
\and P.~Casenove\orcid{0009-0006-6736-1670}\inst{\ref{aff15}}
\and R.~Gimenez\inst{\ref{aff15}}
\and G.~Libet\inst{\ref{aff15}}
\and M.~Ruffenach\orcid{0000-0001-8772-7330}\inst{\ref{aff15}}
\and N.~Aghanim\orcid{0000-0002-6688-8992}\inst{\ref{aff16}}
\and B.~Altieri\orcid{0000-0003-3936-0284}\inst{\ref{aff17}}
\and A.~Amara\inst{\ref{aff18}}
\and S.~Andreon\orcid{0000-0002-2041-8784}\inst{\ref{aff10}}
\and N.~Auricchio\orcid{0000-0003-4444-8651}\inst{\ref{aff4}}
\and H.~Aussel\orcid{0000-0002-1371-5705}\inst{\ref{aff19}}
\and C.~Baccigalupi\orcid{0000-0002-8211-1630}\inst{\ref{aff20},\ref{aff21},\ref{aff22},\ref{aff23}}
\and M.~Baldi\orcid{0000-0003-4145-1943}\inst{\ref{aff5},\ref{aff4},\ref{aff24}}
\and A.~Balestra\orcid{0000-0002-6967-261X}\inst{\ref{aff25}}
\and S.~Bardelli\orcid{0000-0002-8900-0298}\inst{\ref{aff4}}
\and P.~Battaglia\orcid{0000-0002-7337-5909}\inst{\ref{aff4}}
\and A.~Biviano\orcid{0000-0002-0857-0732}\inst{\ref{aff21},\ref{aff20}}
\and A.~Bonchi\orcid{0000-0002-2667-5482}\inst{\ref{aff26}}
\and D.~Bonino\orcid{0000-0002-3336-9977}\inst{\ref{aff27}}
\and E.~Branchini\orcid{0000-0002-0808-6908}\inst{\ref{aff28},\ref{aff29},\ref{aff10}}
\and M.~Brescia\orcid{0000-0001-9506-5680}\inst{\ref{aff30},\ref{aff31}}
\and J.~Brinchmann\orcid{0000-0003-4359-8797}\inst{\ref{aff32},\ref{aff33}}
\and A.~Caillat\inst{\ref{aff1}}
\and S.~Camera\orcid{0000-0003-3399-3574}\inst{\ref{aff34},\ref{aff35},\ref{aff27}}
\and G.~Ca\~nas-Herrera\orcid{0000-0003-2796-2149}\inst{\ref{aff36},\ref{aff37},\ref{aff38}}
\and V.~Capobianco\orcid{0000-0002-3309-7692}\inst{\ref{aff27}}
\and C.~Carbone\orcid{0000-0003-0125-3563}\inst{\ref{aff39}}
\and J.~Carretero\orcid{0000-0002-3130-0204}\inst{\ref{aff40},\ref{aff41}}
\and S.~Casas\orcid{0000-0002-4751-5138}\inst{\ref{aff42}}
\and F.~J.~Castander\orcid{0000-0001-7316-4573}\inst{\ref{aff43},\ref{aff44}}
\and G.~Castignani\orcid{0000-0001-6831-0687}\inst{\ref{aff4}}
\and S.~Cavuoti\orcid{0000-0002-3787-4196}\inst{\ref{aff31},\ref{aff45}}
\and K.~C.~Chambers\orcid{0000-0001-6965-7789}\inst{\ref{aff46}}
\and A.~Cimatti\inst{\ref{aff47}}
\and C.~Colodro-Conde\inst{\ref{aff48}}
\and G.~Congedo\orcid{0000-0003-2508-0046}\inst{\ref{aff13}}
\and C.~J.~Conselice\orcid{0000-0003-1949-7638}\inst{\ref{aff49}}
\and L.~Conversi\orcid{0000-0002-6710-8476}\inst{\ref{aff50},\ref{aff17}}
\and A.~Costille\inst{\ref{aff1}}
\and F.~Courbin\orcid{0000-0003-0758-6510}\inst{\ref{aff51},\ref{aff52}}
\and J.-G.~Cuby\orcid{0000-0002-8767-1442}\inst{\ref{aff53},\ref{aff1}}
\and A.~Da~Silva\orcid{0000-0002-6385-1609}\inst{\ref{aff54},\ref{aff55}}
\and H.~Degaudenzi\orcid{0000-0002-5887-6799}\inst{\ref{aff56}}
\and S.~de~la~Torre\inst{\ref{aff1}}
\and G.~De~Lucia\orcid{0000-0002-6220-9104}\inst{\ref{aff21}}
\and A.~M.~Di~Giorgio\orcid{0000-0002-4767-2360}\inst{\ref{aff57}}
\and H.~Dole\orcid{0000-0002-9767-3839}\inst{\ref{aff16}}
\and M.~Douspis\orcid{0000-0003-4203-3954}\inst{\ref{aff16}}
\and F.~Dubath\orcid{0000-0002-6533-2810}\inst{\ref{aff56}}
\and X.~Dupac\inst{\ref{aff17}}
\and S.~Dusini\orcid{0000-0002-1128-0664}\inst{\ref{aff58}}
\and A.~Ealet\orcid{0000-0003-3070-014X}\inst{\ref{aff6}}
\and S.~Escoffier\orcid{0000-0002-2847-7498}\inst{\ref{aff59}}
\and M.~Fabricius\orcid{0000-0002-7025-6058}\inst{\ref{aff60},\ref{aff61}}
\and M.~Farina\orcid{0000-0002-3089-7846}\inst{\ref{aff57}}
\and R.~Farinelli\inst{\ref{aff4}}
\and F.~Faustini\orcid{0000-0001-6274-5145}\inst{\ref{aff62},\ref{aff26}}
\and S.~Ferriol\inst{\ref{aff6}}
\and S.~Fotopoulou\orcid{0000-0002-9686-254X}\inst{\ref{aff63}}
\and N.~Fourmanoit\orcid{0009-0005-6816-6925}\inst{\ref{aff59}}
\and M.~Frailis\orcid{0000-0002-7400-2135}\inst{\ref{aff21}}
\and E.~Franceschi\orcid{0000-0002-0585-6591}\inst{\ref{aff4}}
\and M.~Fumana\orcid{0000-0001-6787-5950}\inst{\ref{aff39}}
\and S.~Galeotta\orcid{0000-0002-3748-5115}\inst{\ref{aff21}}
\and K.~George\orcid{0000-0002-1734-8455}\inst{\ref{aff61}}
\and W.~Gillard\orcid{0000-0003-4744-9748}\inst{\ref{aff59}}
\and B.~Gillis\orcid{0000-0002-4478-1270}\inst{\ref{aff13}}
\and C.~Giocoli\orcid{0000-0002-9590-7961}\inst{\ref{aff4},\ref{aff24}}
\and J.~Gracia-Carpio\inst{\ref{aff60}}
\and B.~R.~Granett\orcid{0000-0003-2694-9284}\inst{\ref{aff10}}
\and A.~Grazian\orcid{0000-0002-5688-0663}\inst{\ref{aff25}}
\and F.~Grupp\inst{\ref{aff60},\ref{aff61}}
\and S.~V.~H.~Haugan\orcid{0000-0001-9648-7260}\inst{\ref{aff64}}
\and J.~Hoar\inst{\ref{aff17}}
\and H.~Hoekstra\orcid{0000-0002-0641-3231}\inst{\ref{aff38}}
\and W.~Holmes\inst{\ref{aff65}}
\and F.~Hormuth\inst{\ref{aff66}}
\and A.~Hornstrup\orcid{0000-0002-3363-0936}\inst{\ref{aff67},\ref{aff68}}
\and P.~Hudelot\inst{\ref{aff69}}
\and K.~Jahnke\orcid{0000-0003-3804-2137}\inst{\ref{aff70}}
\and M.~Jhabvala\inst{\ref{aff71}}
\and B.~Joachimi\orcid{0000-0001-7494-1303}\inst{\ref{aff72}}
\and E.~Keih\"anen\orcid{0000-0003-1804-7715}\inst{\ref{aff73}}
\and S.~Kermiche\orcid{0000-0002-0302-5735}\inst{\ref{aff59}}
\and A.~Kiessling\orcid{0000-0002-2590-1273}\inst{\ref{aff65}}
\and M.~K\"ummel\orcid{0000-0003-2791-2117}\inst{\ref{aff61}}
\and M.~Kunz\orcid{0000-0002-3052-7394}\inst{\ref{aff74}}
\and H.~Kurki-Suonio\orcid{0000-0002-4618-3063}\inst{\ref{aff75},\ref{aff76}}
\and Q.~Le~Boulc'h\inst{\ref{aff77}}
\and D.~Le~Mignant\orcid{0000-0002-5339-5515}\inst{\ref{aff1}}
\and S.~Ligori\orcid{0000-0003-4172-4606}\inst{\ref{aff27}}
\and P.~B.~Lilje\orcid{0000-0003-4324-7794}\inst{\ref{aff64}}
\and V.~Lindholm\orcid{0000-0003-2317-5471}\inst{\ref{aff75},\ref{aff76}}
\and I.~Lloro\orcid{0000-0001-5966-1434}\inst{\ref{aff78}}
\and G.~Mainetti\orcid{0000-0003-2384-2377}\inst{\ref{aff77}}
\and D.~Maino\inst{\ref{aff9},\ref{aff39},\ref{aff11}}
\and O.~Mansutti\orcid{0000-0001-5758-4658}\inst{\ref{aff21}}
\and S.~Marcin\inst{\ref{aff79}}
\and O.~Marggraf\orcid{0000-0001-7242-3852}\inst{\ref{aff80}}
\and M.~Martinelli\orcid{0000-0002-6943-7732}\inst{\ref{aff62},\ref{aff81}}
\and N.~Martinet\orcid{0000-0003-2786-7790}\inst{\ref{aff1}}
\and F.~Marulli\orcid{0000-0002-8850-0303}\inst{\ref{aff3},\ref{aff4},\ref{aff24}}
\and R.~Massey\orcid{0000-0002-6085-3780}\inst{\ref{aff82}}
\and S.~Maurogordato\inst{\ref{aff83}}
\and E.~Medinaceli\orcid{0000-0002-4040-7783}\inst{\ref{aff4}}
\and S.~Mei\orcid{0000-0002-2849-559X}\inst{\ref{aff84},\ref{aff85}}
\and M.~Melchior\inst{\ref{aff86}}
\and Y.~Mellier\inst{\ref{aff87},\ref{aff69}}
\and M.~Meneghetti\orcid{0000-0003-1225-7084}\inst{\ref{aff4},\ref{aff24}}
\and E.~Merlin\orcid{0000-0001-6870-8900}\inst{\ref{aff62}}
\and G.~Meylan\inst{\ref{aff88}}
\and A.~Mora\orcid{0000-0002-1922-8529}\inst{\ref{aff89}}
\and L.~Moscardini\orcid{0000-0002-3473-6716}\inst{\ref{aff3},\ref{aff4},\ref{aff24}}
\and R.~Nakajima\orcid{0009-0009-1213-7040}\inst{\ref{aff80}}
\and C.~Neissner\orcid{0000-0001-8524-4968}\inst{\ref{aff90},\ref{aff41}}
\and R.~C.~Nichol\orcid{0000-0003-0939-6518}\inst{\ref{aff18}}
\and S.-M.~Niemi\inst{\ref{aff36}}
\and J.~W.~Nightingale\orcid{0000-0002-8987-7401}\inst{\ref{aff91}}
\and C.~Padilla\orcid{0000-0001-7951-0166}\inst{\ref{aff90}}
\and S.~Paltani\orcid{0000-0002-8108-9179}\inst{\ref{aff56}}
\and F.~Pasian\orcid{0000-0002-4869-3227}\inst{\ref{aff21}}
\and K.~Pedersen\inst{\ref{aff92}}
\and W.~J.~Percival\orcid{0000-0002-0644-5727}\inst{\ref{aff93},\ref{aff94},\ref{aff95}}
\and V.~Pettorino\inst{\ref{aff36}}
\and S.~Pires\orcid{0000-0002-0249-2104}\inst{\ref{aff19}}
\and G.~Polenta\orcid{0000-0003-4067-9196}\inst{\ref{aff26}}
\and M.~Poncet\inst{\ref{aff15}}
\and L.~A.~Popa\inst{\ref{aff96}}
\and F.~Raison\orcid{0000-0002-7819-6918}\inst{\ref{aff60}}
\and R.~Rebolo\orcid{0000-0003-3767-7085}\inst{\ref{aff48},\ref{aff97},\ref{aff98}}
\and A.~Renzi\orcid{0000-0001-9856-1970}\inst{\ref{aff99},\ref{aff58}}
\and J.~Rhodes\orcid{0000-0002-4485-8549}\inst{\ref{aff65}}
\and G.~Riccio\inst{\ref{aff31}}
\and E.~Romelli\orcid{0000-0003-3069-9222}\inst{\ref{aff21}}
\and M.~Roncarelli\orcid{0000-0001-9587-7822}\inst{\ref{aff4}}
\and R.~Saglia\orcid{0000-0003-0378-7032}\inst{\ref{aff61},\ref{aff60}}
\and Z.~Sakr\orcid{0000-0002-4823-3757}\inst{\ref{aff100},\ref{aff101},\ref{aff102}}
\and D.~Sapone\orcid{0000-0001-7089-4503}\inst{\ref{aff103}}
\and B.~Sartoris\orcid{0000-0003-1337-5269}\inst{\ref{aff61},\ref{aff21}}
\and M.~Sauvage\orcid{0000-0002-0809-2574}\inst{\ref{aff19}}
\and J.~A.~Schewtschenko\orcid{0000-0002-4913-6393}\inst{\ref{aff13}}
\and M.~Schirmer\orcid{0000-0003-2568-9994}\inst{\ref{aff70}}
\and P.~Schneider\orcid{0000-0001-8561-2679}\inst{\ref{aff80}}
\and T.~Schrabback\orcid{0000-0002-6987-7834}\inst{\ref{aff104}}
\and M.~Scodeggio\inst{\ref{aff39}}
\and A.~Secroun\orcid{0000-0003-0505-3710}\inst{\ref{aff59}}
\and G.~Seidel\orcid{0000-0003-2907-353X}\inst{\ref{aff70}}
\and M.~Seiffert\orcid{0000-0002-7536-9393}\inst{\ref{aff65}}
\and C.~Sirignano\orcid{0000-0002-0995-7146}\inst{\ref{aff99},\ref{aff58}}
\and G.~Sirri\orcid{0000-0003-2626-2853}\inst{\ref{aff24}}
\and L.~Stanco\orcid{0000-0002-9706-5104}\inst{\ref{aff58}}
\and J.~Steinwagner\orcid{0000-0001-7443-1047}\inst{\ref{aff60}}
\and P.~Tallada-Cresp\'{i}\orcid{0000-0002-1336-8328}\inst{\ref{aff40},\ref{aff41}}
\and A.~N.~Taylor\inst{\ref{aff13}}
\and H.~I.~Teplitz\orcid{0000-0002-7064-5424}\inst{\ref{aff105}}
\and I.~Tereno\inst{\ref{aff54},\ref{aff106}}
\and N.~Tessore\orcid{0000-0002-9696-7931}\inst{\ref{aff72}}
\and S.~Toft\orcid{0000-0003-3631-7176}\inst{\ref{aff107},\ref{aff108}}
\and R.~Toledo-Moreo\orcid{0000-0002-2997-4859}\inst{\ref{aff109}}
\and F.~Torradeflot\orcid{0000-0003-1160-1517}\inst{\ref{aff41},\ref{aff40}}
\and I.~Tutusaus\orcid{0000-0002-3199-0399}\inst{\ref{aff101}}
\and L.~Valenziano\orcid{0000-0002-1170-0104}\inst{\ref{aff4},\ref{aff110}}
\and J.~Valiviita\orcid{0000-0001-6225-3693}\inst{\ref{aff75},\ref{aff76}}
\and T.~Vassallo\orcid{0000-0001-6512-6358}\inst{\ref{aff61},\ref{aff21}}
\and G.~Verdoes~Kleijn\orcid{0000-0001-5803-2580}\inst{\ref{aff111}}
\and A.~Veropalumbo\orcid{0000-0003-2387-1194}\inst{\ref{aff10},\ref{aff29},\ref{aff28}}
\and Y.~Wang\orcid{0000-0002-4749-2984}\inst{\ref{aff105}}
\and J.~Weller\orcid{0000-0002-8282-2010}\inst{\ref{aff61},\ref{aff60}}
\and A.~Zacchei\orcid{0000-0003-0396-1192}\inst{\ref{aff21},\ref{aff20}}
\and F.~M.~Zerbi\inst{\ref{aff10}}
\and I.~A.~Zinchenko\orcid{0000-0002-2944-2449}\inst{\ref{aff61}}
\and E.~Zucca\orcid{0000-0002-5845-8132}\inst{\ref{aff4}}
\and V.~Allevato\orcid{0000-0001-7232-5152}\inst{\ref{aff31}}
\and M.~Ballardini\orcid{0000-0003-4481-3559}\inst{\ref{aff112},\ref{aff113},\ref{aff4}}
\and M.~Bolzonella\orcid{0000-0003-3278-4607}\inst{\ref{aff4}}
\and E.~Bozzo\orcid{0000-0002-8201-1525}\inst{\ref{aff56}}
\and C.~Burigana\orcid{0000-0002-3005-5796}\inst{\ref{aff114},\ref{aff110}}
\and R.~Cabanac\orcid{0000-0001-6679-2600}\inst{\ref{aff101}}
\and A.~Cappi\inst{\ref{aff4},\ref{aff83}}
\and D.~Di~Ferdinando\inst{\ref{aff24}}
\and J.~A.~Escartin~Vigo\inst{\ref{aff60}}
\and G.~Fabbian\orcid{0000-0002-3255-4695}\inst{\ref{aff115}}
\and L.~Gabarra\orcid{0000-0002-8486-8856}\inst{\ref{aff116}}
\and W.~G.~Hartley\inst{\ref{aff56}}
\and J.~Mart\'{i}n-Fleitas\orcid{0000-0002-8594-569X}\inst{\ref{aff89}}
\and S.~Matthew\orcid{0000-0001-8448-1697}\inst{\ref{aff13}}
\and M.~Maturi\orcid{0000-0002-3517-2422}\inst{\ref{aff100},\ref{aff117}}
\and N.~Mauri\orcid{0000-0001-8196-1548}\inst{\ref{aff47},\ref{aff24}}
\and R.~B.~Metcalf\orcid{0000-0003-3167-2574}\inst{\ref{aff3},\ref{aff4}}
\and A.~Pezzotta\orcid{0000-0003-0726-2268}\inst{\ref{aff118},\ref{aff60}}
\and M.~P\"ontinen\orcid{0000-0001-5442-2530}\inst{\ref{aff75}}
\and C.~Porciani\orcid{0000-0002-7797-2508}\inst{\ref{aff80}}
\and I.~Risso\orcid{0000-0003-2525-7761}\inst{\ref{aff119}}
\and V.~Scottez\inst{\ref{aff87},\ref{aff120}}
\and M.~Sereno\orcid{0000-0003-0302-0325}\inst{\ref{aff4},\ref{aff24}}
\and M.~Tenti\orcid{0000-0002-4254-5901}\inst{\ref{aff24}}
\and M.~Viel\orcid{0000-0002-2642-5707}\inst{\ref{aff20},\ref{aff21},\ref{aff23},\ref{aff22},\ref{aff121}}
\and M.~Wiesmann\orcid{0009-0000-8199-5860}\inst{\ref{aff64}}
\and Y.~Akrami\orcid{0000-0002-2407-7956}\inst{\ref{aff122},\ref{aff123}}
\and S.~Alvi\orcid{0000-0001-5779-8568}\inst{\ref{aff112}}
\and I.~T.~Andika\orcid{0000-0001-6102-9526}\inst{\ref{aff124},\ref{aff125}}
\and S.~Anselmi\orcid{0000-0002-3579-9583}\inst{\ref{aff58},\ref{aff99},\ref{aff126}}
\and M.~Archidiacono\orcid{0000-0003-4952-9012}\inst{\ref{aff9},\ref{aff11}}
\and F.~Atrio-Barandela\orcid{0000-0002-2130-2513}\inst{\ref{aff127}}
\and S.~Avila\orcid{0000-0001-5043-3662}\inst{\ref{aff40}}
\and M.~Bella\orcid{0000-0002-6406-4789}\inst{\ref{aff101}}
\and P.~Bergamini\orcid{0000-0003-1383-9414}\inst{\ref{aff9},\ref{aff4}}
\and D.~Bertacca\orcid{0000-0002-2490-7139}\inst{\ref{aff99},\ref{aff25},\ref{aff58}}
\and L.~Blot\orcid{0000-0002-9622-7167}\inst{\ref{aff128},\ref{aff12}}
\and S.~Borgani\orcid{0000-0001-6151-6439}\inst{\ref{aff129},\ref{aff20},\ref{aff21},\ref{aff22},\ref{aff121}}
\and M.~L.~Brown\orcid{0000-0002-0370-8077}\inst{\ref{aff49}}
\and S.~Bruton\orcid{0000-0002-6503-5218}\inst{\ref{aff130}}
\and A.~Calabro\orcid{0000-0003-2536-1614}\inst{\ref{aff62}}
\and B.~Camacho~Quevedo\orcid{0000-0002-8789-4232}\inst{\ref{aff44},\ref{aff43}}
\and F.~Caro\inst{\ref{aff62}}
\and C.~S.~Carvalho\inst{\ref{aff106}}
\and T.~Castro\orcid{0000-0002-6292-3228}\inst{\ref{aff21},\ref{aff22},\ref{aff20},\ref{aff121}}
\and Y.~Charles\inst{\ref{aff1}}
\and R.~Chary\orcid{0000-0001-7583-0621}\inst{\ref{aff105},\ref{aff131}}
\and F.~Cogato\orcid{0000-0003-4632-6113}\inst{\ref{aff3},\ref{aff4}}
\and A.~R.~Cooray\orcid{0000-0002-3892-0190}\inst{\ref{aff132}}
\and O.~Cucciati\orcid{0000-0002-9336-7551}\inst{\ref{aff4}}
\and S.~Davini\orcid{0000-0003-3269-1718}\inst{\ref{aff29}}
\and F.~De~Paolis\orcid{0000-0001-6460-7563}\inst{\ref{aff133},\ref{aff134},\ref{aff135}}
\and G.~Desprez\orcid{0000-0001-8325-1742}\inst{\ref{aff111}}
\and A.~D\'iaz-S\'anchez\orcid{0000-0003-0748-4768}\inst{\ref{aff136}}
\and J.~J.~Diaz\inst{\ref{aff48}}
\and S.~Di~Domizio\orcid{0000-0003-2863-5895}\inst{\ref{aff28},\ref{aff29}}
\and J.~M.~Diego\orcid{0000-0001-9065-3926}\inst{\ref{aff137}}
\and P.~Dimauro\orcid{0000-0001-7399-2854}\inst{\ref{aff62},\ref{aff138}}
\and P.-A.~Duc\orcid{0000-0003-3343-6284}\inst{\ref{aff2}}
\and Y.~Fang\inst{\ref{aff61}}
\and A.~M.~N.~Ferguson\inst{\ref{aff13}}
\and A.~G.~Ferrari\orcid{0009-0005-5266-4110}\inst{\ref{aff24}}
\and A.~Finoguenov\orcid{0000-0002-4606-5403}\inst{\ref{aff75}}
\and A.~Fontana\orcid{0000-0003-3820-2823}\inst{\ref{aff62}}
\and A.~Franco\orcid{0000-0002-4761-366X}\inst{\ref{aff134},\ref{aff133},\ref{aff135}}
\and K.~Ganga\orcid{0000-0001-8159-8208}\inst{\ref{aff84}}
\and J.~Garc\'ia-Bellido\orcid{0000-0002-9370-8360}\inst{\ref{aff122}}
\and T.~Gasparetto\orcid{0000-0002-7913-4866}\inst{\ref{aff21}}
\and V.~Gautard\inst{\ref{aff139}}
\and E.~Gaztanaga\orcid{0000-0001-9632-0815}\inst{\ref{aff43},\ref{aff44},\ref{aff140}}
\and F.~Giacomini\orcid{0000-0002-3129-2814}\inst{\ref{aff24}}
\and F.~Gianotti\orcid{0000-0003-4666-119X}\inst{\ref{aff4}}
\and G.~Gozaliasl\orcid{0000-0002-0236-919X}\inst{\ref{aff141},\ref{aff75}}
\and A.~Gregorio\orcid{0000-0003-4028-8785}\inst{\ref{aff129},\ref{aff21},\ref{aff22}}
\and M.~Guidi\orcid{0000-0001-9408-1101}\inst{\ref{aff5},\ref{aff4}}
\and C.~M.~Gutierrez\orcid{0000-0001-7854-783X}\inst{\ref{aff142}}
\and A.~Hall\orcid{0000-0002-3139-8651}\inst{\ref{aff13}}
\and C.~Hern\'andez-Monteagudo\orcid{0000-0001-5471-9166}\inst{\ref{aff98},\ref{aff48}}
\and H.~Hildebrandt\orcid{0000-0002-9814-3338}\inst{\ref{aff143}}
\and J.~Hjorth\orcid{0000-0002-4571-2306}\inst{\ref{aff92}}
\and J.~J.~E.~Kajava\orcid{0000-0002-3010-8333}\inst{\ref{aff144},\ref{aff145}}
\and Y.~Kang\orcid{0009-0000-8588-7250}\inst{\ref{aff56}}
\and V.~Kansal\orcid{0000-0002-4008-6078}\inst{\ref{aff146},\ref{aff147}}
\and D.~Karagiannis\orcid{0000-0002-4927-0816}\inst{\ref{aff112},\ref{aff148}}
\and K.~Kiiveri\inst{\ref{aff73}}
\and C.~C.~Kirkpatrick\inst{\ref{aff73}}
\and S.~Kruk\orcid{0000-0001-8010-8879}\inst{\ref{aff17}}
\and L.~Legrand\orcid{0000-0003-0610-5252}\inst{\ref{aff149},\ref{aff150}}
\and M.~Lembo\orcid{0000-0002-5271-5070}\inst{\ref{aff112},\ref{aff113}}
\and F.~Lepori\orcid{0009-0000-5061-7138}\inst{\ref{aff151}}
\and G.~F.~Lesci\orcid{0000-0002-4607-2830}\inst{\ref{aff3},\ref{aff4}}
\and J.~Lesgourgues\orcid{0000-0001-7627-353X}\inst{\ref{aff42}}
\and L.~Leuzzi\orcid{0009-0006-4479-7017}\inst{\ref{aff3},\ref{aff4}}
\and T.~I.~Liaudat\orcid{0000-0002-9104-314X}\inst{\ref{aff152}}
\and S.~J.~Liu\orcid{0000-0001-7680-2139}\inst{\ref{aff57}}
\and A.~Loureiro\orcid{0000-0002-4371-0876}\inst{\ref{aff153},\ref{aff154}}
\and J.~Macias-Perez\orcid{0000-0002-5385-2763}\inst{\ref{aff155}}
\and M.~Magliocchetti\orcid{0000-0001-9158-4838}\inst{\ref{aff57}}
\and E.~A.~Magnier\orcid{0000-0002-7965-2815}\inst{\ref{aff46}}
\and C.~Mancini\orcid{0000-0002-4297-0561}\inst{\ref{aff39}}
\and F.~Mannucci\orcid{0000-0002-4803-2381}\inst{\ref{aff156}}
\and R.~Maoli\orcid{0000-0002-6065-3025}\inst{\ref{aff157},\ref{aff62}}
\and C.~J.~A.~P.~Martins\orcid{0000-0002-4886-9261}\inst{\ref{aff158},\ref{aff32}}
\and L.~Maurin\orcid{0000-0002-8406-0857}\inst{\ref{aff16}}
\and M.~Miluzio\inst{\ref{aff17},\ref{aff159}}
\and P.~Monaco\orcid{0000-0003-2083-7564}\inst{\ref{aff129},\ref{aff21},\ref{aff22},\ref{aff20}}
\and A.~Montoro\orcid{0000-0003-4730-8590}\inst{\ref{aff43},\ref{aff44}}
\and C.~Moretti\orcid{0000-0003-3314-8936}\inst{\ref{aff23},\ref{aff121},\ref{aff21},\ref{aff20},\ref{aff22}}
\and G.~Morgante\inst{\ref{aff4}}
\and S.~Nadathur\orcid{0000-0001-9070-3102}\inst{\ref{aff140}}
\and K.~Naidoo\orcid{0000-0002-9182-1802}\inst{\ref{aff140}}
\and A.~Navarro-Alsina\orcid{0000-0002-3173-2592}\inst{\ref{aff80}}
\and S.~Nesseris\orcid{0000-0002-0567-0324}\inst{\ref{aff122}}
\and F.~Passalacqua\orcid{0000-0002-8606-4093}\inst{\ref{aff99},\ref{aff58}}
\and K.~Paterson\orcid{0000-0001-8340-3486}\inst{\ref{aff70}}
\and L.~Patrizii\inst{\ref{aff24}}
\and A.~Pisani\orcid{0000-0002-6146-4437}\inst{\ref{aff59}}
\and D.~Potter\orcid{0000-0002-0757-5195}\inst{\ref{aff151}}
\and M.~Radovich\orcid{0000-0002-3585-866X}\inst{\ref{aff25}}
\and P.-F.~Rocci\inst{\ref{aff16}}
\and S.~Sacquegna\orcid{0000-0002-8433-6630}\inst{\ref{aff133},\ref{aff134},\ref{aff135}}
\and M.~Sahl\'en\orcid{0000-0003-0973-4804}\inst{\ref{aff160}}
\and D.~B.~Sanders\orcid{0000-0002-1233-9998}\inst{\ref{aff46}}
\and E.~Sarpa\orcid{0000-0002-1256-655X}\inst{\ref{aff23},\ref{aff121},\ref{aff22}}
\and A.~Schneider\orcid{0000-0001-7055-8104}\inst{\ref{aff151}}
\and D.~Sciotti\orcid{0009-0008-4519-2620}\inst{\ref{aff62},\ref{aff81}}
\and E.~Sellentin\inst{\ref{aff161},\ref{aff38}}
\and F.~Shankar\orcid{0000-0001-8973-5051}\inst{\ref{aff162}}
\and L.~C.~Smith\orcid{0000-0002-3259-2771}\inst{\ref{aff163}}
\and K.~Tanidis\orcid{0000-0001-9843-5130}\inst{\ref{aff116}}
\and G.~Testera\inst{\ref{aff29}}
\and R.~Teyssier\orcid{0000-0001-7689-0933}\inst{\ref{aff164}}
\and S.~Tosi\orcid{0000-0002-7275-9193}\inst{\ref{aff28},\ref{aff29},\ref{aff10}}
\and A.~Troja\orcid{0000-0003-0239-4595}\inst{\ref{aff99},\ref{aff58}}
\and M.~Tucci\inst{\ref{aff56}}
\and C.~Valieri\inst{\ref{aff24}}
\and A.~Venhola\orcid{0000-0001-6071-4564}\inst{\ref{aff165}}
\and G.~Verza\orcid{0000-0002-1886-8348}\inst{\ref{aff166}}
\and P.~Vielzeuf\orcid{0000-0003-2035-9339}\inst{\ref{aff59}}
\and N.~A.~Walton\orcid{0000-0003-3983-8778}\inst{\ref{aff163}}
\and J.~R.~Weaver\orcid{0000-0003-1614-196X}\inst{\ref{aff167}}
\and L.~Zalesky\orcid{0000-0001-5680-2326}\inst{\ref{aff46}}
\and J.~G.~Sorce\orcid{0000-0002-2307-2432}\inst{\ref{aff168},\ref{aff16}}}
										   
\institute{Aix-Marseille Universit\'e, CNRS, CNES, LAM, Marseille, France\label{aff1}
\and
Universit\'e de Strasbourg, CNRS, Observatoire astronomique de Strasbourg, UMR 7550, 67000 Strasbourg, France\label{aff2}
\and
Dipartimento di Fisica e Astronomia "Augusto Righi" - Alma Mater Studiorum Universit\`a di Bologna, via Piero Gobetti 93/2, 40129 Bologna, Italy\label{aff3}
\and
INAF-Osservatorio di Astrofisica e Scienza dello Spazio di Bologna, Via Piero Gobetti 93/3, 40129 Bologna, Italy\label{aff4}
\and
Dipartimento di Fisica e Astronomia, Universit\`a di Bologna, Via Gobetti 93/2, 40129 Bologna, Italy\label{aff5}
\and
Universit\'e Claude Bernard Lyon 1, CNRS/IN2P3, IP2I Lyon, UMR 5822, Villeurbanne, F-69100, France\label{aff6}
\and
Minnesota Institute for Astrophysics, University of Minnesota, 116 Church St SE, Minneapolis, MN 55455, USA\label{aff7}
\and
UCB Lyon 1, CNRS/IN2P3, IUF, IP2I Lyon, 4 rue Enrico Fermi, 69622 Villeurbanne, France\label{aff8}
\and
Dipartimento di Fisica "Aldo Pontremoli", Universit\`a degli Studi di Milano, Via Celoria 16, 20133 Milano, Italy\label{aff9}
\and
INAF-Osservatorio Astronomico di Brera, Via Brera 28, 20122 Milano, Italy\label{aff10}
\and
INFN-Sezione di Milano, Via Celoria 16, 20133 Milano, Italy\label{aff11}
\and
Laboratoire d'etude de l'Univers et des phenomenes eXtremes, Observatoire de Paris, Universit\'e PSL, Sorbonne Universit\'e, CNRS, 92190 Meudon, France\label{aff12}
\and
Institute for Astronomy, University of Edinburgh, Royal Observatory, Blackford Hill, Edinburgh EH9 3HJ, UK\label{aff13}
\and
Department of Physics and Astronomy, University of British Columbia, Vancouver, BC V6T 1Z1, Canada\label{aff14}
\and
Centre National d'Etudes Spatiales -- Centre spatial de Toulouse, 18 avenue Edouard Belin, 31401 Toulouse Cedex 9, France\label{aff15}
\and
Universit\'e Paris-Saclay, CNRS, Institut d'astrophysique spatiale, 91405, Orsay, France\label{aff16}
\and
ESAC/ESA, Camino Bajo del Castillo, s/n., Urb. Villafranca del Castillo, 28692 Villanueva de la Ca\~nada, Madrid, Spain\label{aff17}
\and
School of Mathematics and Physics, University of Surrey, Guildford, Surrey, GU2 7XH, UK\label{aff18}
\and
Universit\'e Paris-Saclay, Universit\'e Paris Cit\'e, CEA, CNRS, AIM, 91191, Gif-sur-Yvette, France\label{aff19}
\and
IFPU, Institute for Fundamental Physics of the Universe, via Beirut 2, 34151 Trieste, Italy\label{aff20}
\and
INAF-Osservatorio Astronomico di Trieste, Via G. B. Tiepolo 11, 34143 Trieste, Italy\label{aff21}
\and
INFN, Sezione di Trieste, Via Valerio 2, 34127 Trieste TS, Italy\label{aff22}
\and
SISSA, International School for Advanced Studies, Via Bonomea 265, 34136 Trieste TS, Italy\label{aff23}
\and
INFN-Sezione di Bologna, Viale Berti Pichat 6/2, 40127 Bologna, Italy\label{aff24}
\and
INAF-Osservatorio Astronomico di Padova, Via dell'Osservatorio 5, 35122 Padova, Italy\label{aff25}
\and
Space Science Data Center, Italian Space Agency, via del Politecnico snc, 00133 Roma, Italy\label{aff26}
\and
INAF-Osservatorio Astrofisico di Torino, Via Osservatorio 20, 10025 Pino Torinese (TO), Italy\label{aff27}
\and
Dipartimento di Fisica, Universit\`a di Genova, Via Dodecaneso 33, 16146, Genova, Italy\label{aff28}
\and
INFN-Sezione di Genova, Via Dodecaneso 33, 16146, Genova, Italy\label{aff29}
\and
Department of Physics "E. Pancini", University Federico II, Via Cinthia 6, 80126, Napoli, Italy\label{aff30}
\and
INAF-Osservatorio Astronomico di Capodimonte, Via Moiariello 16, 80131 Napoli, Italy\label{aff31}
\and
Instituto de Astrof\'isica e Ci\^encias do Espa\c{c}o, Universidade do Porto, CAUP, Rua das Estrelas, PT4150-762 Porto, Portugal\label{aff32}
\and
Faculdade de Ci\^encias da Universidade do Porto, Rua do Campo de Alegre, 4150-007 Porto, Portugal\label{aff33}
\and
Dipartimento di Fisica, Universit\`a degli Studi di Torino, Via P. Giuria 1, 10125 Torino, Italy\label{aff34}
\and
INFN-Sezione di Torino, Via P. Giuria 1, 10125 Torino, Italy\label{aff35}
\and
European Space Agency/ESTEC, Keplerlaan 1, 2201 AZ Noordwijk, The Netherlands\label{aff36}
\and
Institute Lorentz, Leiden University, Niels Bohrweg 2, 2333 CA Leiden, The Netherlands\label{aff37}
\and
Leiden Observatory, Leiden University, Einsteinweg 55, 2333 CC Leiden, The Netherlands\label{aff38}
\and
INAF-IASF Milano, Via Alfonso Corti 12, 20133 Milano, Italy\label{aff39}
\and
Centro de Investigaciones Energ\'eticas, Medioambientales y Tecnol\'ogicas (CIEMAT), Avenida Complutense 40, 28040 Madrid, Spain\label{aff40}
\and
Port d'Informaci\'{o} Cient\'{i}fica, Campus UAB, C. Albareda s/n, 08193 Bellaterra (Barcelona), Spain\label{aff41}
\and
Institute for Theoretical Particle Physics and Cosmology (TTK), RWTH Aachen University, 52056 Aachen, Germany\label{aff42}
\and
Institute of Space Sciences (ICE, CSIC), Campus UAB, Carrer de Can Magrans, s/n, 08193 Barcelona, Spain\label{aff43}
\and
Institut d'Estudis Espacials de Catalunya (IEEC),  Edifici RDIT, Campus UPC, 08860 Castelldefels, Barcelona, Spain\label{aff44}
\and
INFN section of Naples, Via Cinthia 6, 80126, Napoli, Italy\label{aff45}
\and
Institute for Astronomy, University of Hawaii, 2680 Woodlawn Drive, Honolulu, HI 96822, USA\label{aff46}
\and
Dipartimento di Fisica e Astronomia "Augusto Righi" - Alma Mater Studiorum Universit\`a di Bologna, Viale Berti Pichat 6/2, 40127 Bologna, Italy\label{aff47}
\and
Instituto de Astrof\'{\i}sica de Canarias, V\'{\i}a L\'actea, 38205 La Laguna, Tenerife, Spain\label{aff48}
\and
Jodrell Bank Centre for Astrophysics, Department of Physics and Astronomy, University of Manchester, Oxford Road, Manchester M13 9PL, UK\label{aff49}
\and
European Space Agency/ESRIN, Largo Galileo Galilei 1, 00044 Frascati, Roma, Italy\label{aff50}
\and
Institut de Ci\`{e}ncies del Cosmos (ICCUB), Universitat de Barcelona (IEEC-UB), Mart\'{i} i Franqu\`{e}s 1, 08028 Barcelona, Spain\label{aff51}
\and
Instituci\'o Catalana de Recerca i Estudis Avan\c{c}ats (ICREA), Passeig de Llu\'{\i}s Companys 23, 08010 Barcelona, Spain\label{aff52}
\and
Canada-France-Hawaii Telescope, 65-1238 Mamalahoa Hwy, Kamuela, HI 96743, USA\label{aff53}
\and
Departamento de F\'isica, Faculdade de Ci\^encias, Universidade de Lisboa, Edif\'icio C8, Campo Grande, PT1749-016 Lisboa, Portugal\label{aff54}
\and
Instituto de Astrof\'isica e Ci\^encias do Espa\c{c}o, Faculdade de Ci\^encias, Universidade de Lisboa, Campo Grande, 1749-016 Lisboa, Portugal\label{aff55}
\and
Department of Astronomy, University of Geneva, ch. d'Ecogia 16, 1290 Versoix, Switzerland\label{aff56}
\and
INAF-Istituto di Astrofisica e Planetologia Spaziali, via del Fosso del Cavaliere, 100, 00100 Roma, Italy\label{aff57}
\and
INFN-Padova, Via Marzolo 8, 35131 Padova, Italy\label{aff58}
\and
Aix-Marseille Universit\'e, CNRS/IN2P3, CPPM, Marseille, France\label{aff59}
\and
Max Planck Institute for Extraterrestrial Physics, Giessenbachstr. 1, 85748 Garching, Germany\label{aff60}
\and
Universit\"ats-Sternwarte M\"unchen, Fakult\"at f\"ur Physik, Ludwig-Maximilians-Universit\"at M\"unchen, Scheinerstrasse 1, 81679 M\"unchen, Germany\label{aff61}
\and
INAF-Osservatorio Astronomico di Roma, Via Frascati 33, 00078 Monteporzio Catone, Italy\label{aff62}
\and
School of Physics, HH Wills Physics Laboratory, University of Bristol, Tyndall Avenue, Bristol, BS8 1TL, UK\label{aff63}
\and
Institute of Theoretical Astrophysics, University of Oslo, P.O. Box 1029 Blindern, 0315 Oslo, Norway\label{aff64}
\and
Jet Propulsion Laboratory, California Institute of Technology, 4800 Oak Grove Drive, Pasadena, CA, 91109, USA\label{aff65}
\and
Felix Hormuth Engineering, Goethestr. 17, 69181 Leimen, Germany\label{aff66}
\and
Technical University of Denmark, Elektrovej 327, 2800 Kgs. Lyngby, Denmark\label{aff67}
\and
Cosmic Dawn Center (DAWN), Denmark\label{aff68}
\and
Institut d'Astrophysique de Paris, UMR 7095, CNRS, and Sorbonne Universit\'e, 98 bis boulevard Arago, 75014 Paris, France\label{aff69}
\and
Max-Planck-Institut f\"ur Astronomie, K\"onigstuhl 17, 69117 Heidelberg, Germany\label{aff70}
\and
NASA Goddard Space Flight Center, Greenbelt, MD 20771, USA\label{aff71}
\and
Department of Physics and Astronomy, University College London, Gower Street, London WC1E 6BT, UK\label{aff72}
\and
Department of Physics and Helsinki Institute of Physics, Gustaf H\"allstr\"omin katu 2, 00014 University of Helsinki, Finland\label{aff73}
\and
Universit\'e de Gen\`eve, D\'epartement de Physique Th\'eorique and Centre for Astroparticle Physics, 24 quai Ernest-Ansermet, CH-1211 Gen\`eve 4, Switzerland\label{aff74}
\and
Department of Physics, P.O. Box 64, 00014 University of Helsinki, Finland\label{aff75}
\and
Helsinki Institute of Physics, Gustaf H{\"a}llstr{\"o}min katu 2, University of Helsinki, Helsinki, Finland\label{aff76}
\and
Centre de Calcul de l'IN2P3/CNRS, 21 avenue Pierre de Coubertin 69627 Villeurbanne Cedex, France\label{aff77}
\and
SKA Observatory, Jodrell Bank, Lower Withington, Macclesfield, Cheshire SK11 9FT, UK\label{aff78}
\and
University of Applied Sciences and Arts of Northwestern Switzerland, School of Computer Science, 5210 Windisch, Switzerland\label{aff79}
\and
Universit\"at Bonn, Argelander-Institut f\"ur Astronomie, Auf dem H\"ugel 71, 53121 Bonn, Germany\label{aff80}
\and
INFN-Sezione di Roma, Piazzale Aldo Moro, 2 - c/o Dipartimento di Fisica, Edificio G. Marconi, 00185 Roma, Italy\label{aff81}
\and
Department of Physics, Institute for Computational Cosmology, Durham University, South Road, Durham, DH1 3LE, UK\label{aff82}
\and
Universit\'e C\^{o}te d'Azur, Observatoire de la C\^{o}te d'Azur, CNRS, Laboratoire Lagrange, Bd de l'Observatoire, CS 34229, 06304 Nice cedex 4, France\label{aff83}
\and
Universit\'e Paris Cit\'e, CNRS, Astroparticule et Cosmologie, 75013 Paris, France\label{aff84}
\and
CNRS-UCB International Research Laboratory, Centre Pierre Binetruy, IRL2007, CPB-IN2P3, Berkeley, USA\label{aff85}
\and
University of Applied Sciences and Arts of Northwestern Switzerland, School of Engineering, 5210 Windisch, Switzerland\label{aff86}
\and
Institut d'Astrophysique de Paris, 98bis Boulevard Arago, 75014, Paris, France\label{aff87}
\and
Institute of Physics, Laboratory of Astrophysics, Ecole Polytechnique F\'ed\'erale de Lausanne (EPFL), Observatoire de Sauverny, 1290 Versoix, Switzerland\label{aff88}
\and
Aurora Technology for European Space Agency (ESA), Camino bajo del Castillo, s/n, Urbanizacion Villafranca del Castillo, Villanueva de la Ca\~nada, 28692 Madrid, Spain\label{aff89}
\and
Institut de F\'{i}sica d'Altes Energies (IFAE), The Barcelona Institute of Science and Technology, Campus UAB, 08193 Bellaterra (Barcelona), Spain\label{aff90}
\and
School of Mathematics, Statistics and Physics, Newcastle University, Herschel Building, Newcastle-upon-Tyne, NE1 7RU, UK\label{aff91}
\and
DARK, Niels Bohr Institute, University of Copenhagen, Jagtvej 155, 2200 Copenhagen, Denmark\label{aff92}
\and
Waterloo Centre for Astrophysics, University of Waterloo, Waterloo, Ontario N2L 3G1, Canada\label{aff93}
\and
Department of Physics and Astronomy, University of Waterloo, Waterloo, Ontario N2L 3G1, Canada\label{aff94}
\and
Perimeter Institute for Theoretical Physics, Waterloo, Ontario N2L 2Y5, Canada\label{aff95}
\and
Institute of Space Science, Str. Atomistilor, nr. 409 M\u{a}gurele, Ilfov, 077125, Romania\label{aff96}
\and
Consejo Superior de Investigaciones Cientificas, Calle Serrano 117, 28006 Madrid, Spain\label{aff97}
\and
Universidad de La Laguna, Departamento de Astrof\'{\i}sica, 38206 La Laguna, Tenerife, Spain\label{aff98}
\and
Dipartimento di Fisica e Astronomia "G. Galilei", Universit\`a di Padova, Via Marzolo 8, 35131 Padova, Italy\label{aff99}
\and
Institut f\"ur Theoretische Physik, University of Heidelberg, Philosophenweg 16, 69120 Heidelberg, Germany\label{aff100}
\and
Institut de Recherche en Astrophysique et Plan\'etologie (IRAP), Universit\'e de Toulouse, CNRS, UPS, CNES, 14 Av. Edouard Belin, 31400 Toulouse, France\label{aff101}
\and
Universit\'e St Joseph; Faculty of Sciences, Beirut, Lebanon\label{aff102}
\and
Departamento de F\'isica, FCFM, Universidad de Chile, Blanco Encalada 2008, Santiago, Chile\label{aff103}
\and
Universit\"at Innsbruck, Institut f\"ur Astro- und Teilchenphysik, Technikerstr. 25/8, 6020 Innsbruck, Austria\label{aff104}
\and
Infrared Processing and Analysis Center, California Institute of Technology, Pasadena, CA 91125, USA\label{aff105}
\and
Instituto de Astrof\'isica e Ci\^encias do Espa\c{c}o, Faculdade de Ci\^encias, Universidade de Lisboa, Tapada da Ajuda, 1349-018 Lisboa, Portugal\label{aff106}
\and
Cosmic Dawn Center (DAWN)\label{aff107}
\and
Niels Bohr Institute, University of Copenhagen, Jagtvej 128, 2200 Copenhagen, Denmark\label{aff108}
\and
Universidad Polit\'ecnica de Cartagena, Departamento de Electr\'onica y Tecnolog\'ia de Computadoras,  Plaza del Hospital 1, 30202 Cartagena, Spain\label{aff109}
\and
INFN-Bologna, Via Irnerio 46, 40126 Bologna, Italy\label{aff110}
\and
Kapteyn Astronomical Institute, University of Groningen, PO Box 800, 9700 AV Groningen, The Netherlands\label{aff111}
\and
Dipartimento di Fisica e Scienze della Terra, Universit\`a degli Studi di Ferrara, Via Giuseppe Saragat 1, 44122 Ferrara, Italy\label{aff112}
\and
Istituto Nazionale di Fisica Nucleare, Sezione di Ferrara, Via Giuseppe Saragat 1, 44122 Ferrara, Italy\label{aff113}
\and
INAF, Istituto di Radioastronomia, Via Piero Gobetti 101, 40129 Bologna, Italy\label{aff114}
\and
School of Physics and Astronomy, Cardiff University, The Parade, Cardiff, CF24 3AA, UK\label{aff115}
\and
Department of Physics, Oxford University, Keble Road, Oxford OX1 3RH, UK\label{aff116}
\and
Zentrum f\"ur Astronomie, Universit\"at Heidelberg, Philosophenweg 12, 69120 Heidelberg, Germany\label{aff117}
\and
INAF - Osservatorio Astronomico di Brera, via Emilio Bianchi 46, 23807 Merate, Italy\label{aff118}
\and
INAF-Osservatorio Astronomico di Brera, Via Brera 28, 20122 Milano, Italy, and INFN-Sezione di Genova, Via Dodecaneso 33, 16146, Genova, Italy\label{aff119}
\and
ICL, Junia, Universit\'e Catholique de Lille, LITL, 59000 Lille, France\label{aff120}
\and
ICSC - Centro Nazionale di Ricerca in High Performance Computing, Big Data e Quantum Computing, Via Magnanelli 2, Bologna, Italy\label{aff121}
\and
Instituto de F\'isica Te\'orica UAM-CSIC, Campus de Cantoblanco, 28049 Madrid, Spain\label{aff122}
\and
CERCA/ISO, Department of Physics, Case Western Reserve University, 10900 Euclid Avenue, Cleveland, OH 44106, USA\label{aff123}
\and
Technical University of Munich, TUM School of Natural Sciences, Physics Department, James-Franck-Str.~1, 85748 Garching, Germany\label{aff124}
\and
Max-Planck-Institut f\"ur Astrophysik, Karl-Schwarzschild-Str.~1, 85748 Garching, Germany\label{aff125}
\and
Laboratoire Univers et Th\'eorie, Observatoire de Paris, Universit\'e PSL, Universit\'e Paris Cit\'e, CNRS, 92190 Meudon, France\label{aff126}
\and
Departamento de F{\'\i}sica Fundamental. Universidad de Salamanca. Plaza de la Merced s/n. 37008 Salamanca, Spain\label{aff127}
\and
Center for Data-Driven Discovery, Kavli IPMU (WPI), UTIAS, The University of Tokyo, Kashiwa, Chiba 277-8583, Japan\label{aff128}
\and
Dipartimento di Fisica - Sezione di Astronomia, Universit\`a di Trieste, Via Tiepolo 11, 34131 Trieste, Italy\label{aff129}
\and
California Institute of Technology, 1200 E California Blvd, Pasadena, CA 91125, USA\label{aff130}
\and
University of California, Los Angeles, CA 90095-1562, USA\label{aff131}
\and
Department of Physics \& Astronomy, University of California Irvine, Irvine CA 92697, USA\label{aff132}
\and
Department of Mathematics and Physics E. De Giorgi, University of Salento, Via per Arnesano, CP-I93, 73100, Lecce, Italy\label{aff133}
\and
INFN, Sezione di Lecce, Via per Arnesano, CP-193, 73100, Lecce, Italy\label{aff134}
\and
INAF-Sezione di Lecce, c/o Dipartimento Matematica e Fisica, Via per Arnesano, 73100, Lecce, Italy\label{aff135}
\and
Departamento F\'isica Aplicada, Universidad Polit\'ecnica de Cartagena, Campus Muralla del Mar, 30202 Cartagena, Murcia, Spain\label{aff136}
\and
Instituto de F\'isica de Cantabria, Edificio Juan Jord\'a, Avenida de los Castros, 39005 Santander, Spain\label{aff137}
\and
Observatorio Nacional, Rua General Jose Cristino, 77-Bairro Imperial de Sao Cristovao, Rio de Janeiro, 20921-400, Brazil\label{aff138}
\and
CEA Saclay, DFR/IRFU, Service d'Astrophysique, Bat. 709, 91191 Gif-sur-Yvette, France\label{aff139}
\and
Institute of Cosmology and Gravitation, University of Portsmouth, Portsmouth PO1 3FX, UK\label{aff140}
\and
Department of Computer Science, Aalto University, PO Box 15400, Espoo, FI-00 076, Finland\label{aff141}
\and
Instituto de Astrof\'\i sica de Canarias, c/ Via Lactea s/n, La Laguna 38200, Spain. Departamento de Astrof\'\i sica de la Universidad de La Laguna, Avda. Francisco Sanchez, La Laguna, 38200, Spain\label{aff142}
\and
Ruhr University Bochum, Faculty of Physics and Astronomy, Astronomical Institute (AIRUB), German Centre for Cosmological Lensing (GCCL), 44780 Bochum, Germany\label{aff143}
\and
Department of Physics and Astronomy, Vesilinnantie 5, 20014 University of Turku, Finland\label{aff144}
\and
Serco for European Space Agency (ESA), Camino bajo del Castillo, s/n, Urbanizacion Villafranca del Castillo, Villanueva de la Ca\~nada, 28692 Madrid, Spain\label{aff145}
\and
ARC Centre of Excellence for Dark Matter Particle Physics, Melbourne, Australia\label{aff146}
\and
Centre for Astrophysics \& Supercomputing, Swinburne University of Technology,  Hawthorn, Victoria 3122, Australia\label{aff147}
\and
Department of Physics and Astronomy, University of the Western Cape, Bellville, Cape Town, 7535, South Africa\label{aff148}
\and
DAMTP, Centre for Mathematical Sciences, Wilberforce Road, Cambridge CB3 0WA, UK\label{aff149}
\and
Kavli Institute for Cosmology Cambridge, Madingley Road, Cambridge, CB3 0HA, UK\label{aff150}
\and
Department of Astrophysics, University of Zurich, Winterthurerstrasse 190, 8057 Zurich, Switzerland\label{aff151}
\and
IRFU, CEA, Universit\'e Paris-Saclay 91191 Gif-sur-Yvette Cedex, France\label{aff152}
\and
Oskar Klein Centre for Cosmoparticle Physics, Department of Physics, Stockholm University, Stockholm, SE-106 91, Sweden\label{aff153}
\and
Astrophysics Group, Blackett Laboratory, Imperial College London, London SW7 2AZ, UK\label{aff154}
\and
Univ. Grenoble Alpes, CNRS, Grenoble INP, LPSC-IN2P3, 53, Avenue des Martyrs, 38000, Grenoble, France\label{aff155}
\and
INAF-Osservatorio Astrofisico di Arcetri, Largo E. Fermi 5, 50125, Firenze, Italy\label{aff156}
\and
Dipartimento di Fisica, Sapienza Universit\`a di Roma, Piazzale Aldo Moro 2, 00185 Roma, Italy\label{aff157}
\and
Centro de Astrof\'{\i}sica da Universidade do Porto, Rua das Estrelas, 4150-762 Porto, Portugal\label{aff158}
\and
HE Space for European Space Agency (ESA), Camino bajo del Castillo, s/n, Urbanizacion Villafranca del Castillo, Villanueva de la Ca\~nada, 28692 Madrid, Spain\label{aff159}
\and
Theoretical astrophysics, Department of Physics and Astronomy, Uppsala University, Box 515, 751 20 Uppsala, Sweden\label{aff160}
\and
Mathematical Institute, University of Leiden, Einsteinweg 55, 2333 CA Leiden, The Netherlands\label{aff161}
\and
School of Physics \& Astronomy, University of Southampton, Highfield Campus, Southampton SO17 1BJ, UK\label{aff162}
\and
Institute of Astronomy, University of Cambridge, Madingley Road, Cambridge CB3 0HA, UK\label{aff163}
\and
Department of Astrophysical Sciences, Peyton Hall, Princeton University, Princeton, NJ 08544, USA\label{aff164}
\and
Space physics and astronomy research unit, University of Oulu, Pentti Kaiteran katu 1, FI-90014 Oulu, Finland\label{aff165}
\and
Center for Computational Astrophysics, Flatiron Institute, 162 5th Avenue, 10010, New York, NY, USA\label{aff166}
\and
Department of Astronomy, University of Massachusetts, Amherst, MA 01003, USA\label{aff167}
\and
Univ. Lille, CNRS, Centrale Lille, UMR 9189 CRIStAL, 59000 Lille, France\label{aff168}}    

\abstract 
{The spectroscopy processing function (SPE PF) of the \Euclid pipeline is dedicated to the automatic analysis of 1D spectra to determine redshifts, line fluxes, and spectral classifications. The first \Euclid Quick Data Release (Q1) delivers these measurements for all $\HE<22.5$ objects identified in the photometric survey. In this paper, we present an overview of the SPE PF algorithm and assess its performance by comparing its results with high-quality spectroscopic redshifts from the Dark Energy Spectroscopic Instrument (DESI) survey in the Euclid Deep Field North. Our findings highlight remarkable accuracy in successful redshift measurements, with a bias of less than $3 \times 10^{-5}$ in $(z_{\rm SPE}-z_{\rm DESI})/(1+z_{\rm DESI})$ and a high precision of approximately $10^{-3}$. The majority of spectra have only a single spectral feature or none at all. To avoid spurious detections, whereby noise features are misinterpreted as lines or lines are misidentified, it is therefore essential to apply well-defined criteria on quantities such as the redshift probability, or the \ha\ flux, and signal-to-noise ratio. Using a well-tuned quality selection, we achieve an 89\% redshift success rate in the target redshift range for cosmology ($0.9<z<1.8$), which is well covered by DESI for $z<1.6$. Outside this range in which the \ha\ line is observable, redshift measurements are less reliable, except for sources showing specific spectral features (e.g. two bright lines or strong continuum). The classification based on spectroscopy alone is effective for galaxies (about 80\% success rate), while it is less efficient for stars and quasars ($<60\%$). Ongoing refinements along the entire chain of PFs are expected to enhance both redshift measurements and spectral classification, allowing us to define the large and reliable sample required for cosmological analyses. Taking into account the planned evolution of the spectroscopic pipeline, partially based on the limitations identified in this paper, these results are encouraging for \Euclid's future galaxy clustering measurements, even though the requirements are not yet fulfilled.}

%
%
%
%
   \titlerunning{\Euclid: Q1 spectroscopic dataset}
   \authorrunning{Euclid Collaboration: V. Le Brun et al.}
   \maketitle
%

%
%
%
   
\section{\label{sc:Intro}Introduction}
One of the two main objectives of the \Euclid mission \citep{EuclidSkyOverview} is to measure galaxy clustering in the redshift interval $0.84<z<1.88$ (hereafter, the `target range'), using spectroscopic redshifts measured on the spectra provided by the Near Infrared Spectro-Photometer (NISP) instrument \citep{EuclidSkyNISP}. The spectra have a resolution of $R\simeq 500$, cover the wavelength range 1206--1892\,\si{\nm}, and redshifts in the target range are measured by identifying the \ha\ emission line. NISP is a slitless spectrometer, which means that the spectra of all objects on the sky potentially leave a trace on the spectroscopic exposures. The only condition for a spectrum to be extracted is that a corresponding object be identified in the photometric catalogue constructed from the \Euclid imaging. This paper presents spectroscopic measurements and redshifts  released in the framework of the first `Quick Release' of \Euclid products \citep{Q1cite,Q1-TP001}, for objects brighter than $\HE=22.5$ and observed over the area of the Euclid Deep Fields (EDF) in the configuration of the Euclid Wide Survey (EWS). However, given the specific objectives of the mission, in terms of redshift range and emission-line flux, as well as wavelength domain and intrinsic quality of the spectra, these data cannot be used as a homogeneous set for blind statistical studies, and special attention must be paid to redshift values outside the target range or with \ha\ lines fainter than the nominal limit of the survey. In addition, we describe the output and performance of the sole spectroscopy channel, while the final analysis of \Euclid will benefit from the combination with the high-quality visible and infrared images provided by both the VIS \citep{EuclidSkyVIS} and NISP instruments.

The paper is organised as follows. Section~\ref{sc:Products} describes the various spectroscopic products that are present in the \Euclid scientific archive, together with the main parameters which are useful in extracting the best measurements, as well as a brief description of the methods of redshift calculation. Section~\ref{sc:perfs} gives results on the efficiency and accuracy of the redshift measurements, while the results of the spectroscopic classification are presented in Sect.~\ref{sc:classification}.
Section~\ref{sc:LineFlux} gives information about the quality and performance of the emission-line fluxes, and Sect.~\ref{sc:Purity} discusses the quality of the redshift measurement for various sub-samples with respect to existing redshift catalogues. 

\section{\label{sc:Products}SPE PF methods and product description}
\subsection{SPE PF spectra handling and basic description}

\label{sc:basic_description}

The \Euclid data processing is handled by the Science Ground Segment, which is split into processing functions (PFs), each taking care of a specific aspect. The spectroscopy PF (SPE PF) is in charge of redshift estimation, measurement-reliability calculation, and spectroscopic classification of all objects present in the main catalogue created by the `merge' PF \citep[MER PF,][]{Q1-TP004}. The SPE PF first reads the 1D spectra produced by the SIR PF \citep{Q1-TP006} by combining the four individual exposures taken with different orientations of the grisms (see \citealp{Scaramella-EP1} for details on the observing strategy). The data provided by the SIR PF also include flags describing the quality of each pixel in the spectrum and the number of individual exposures that actually contributed to its final value. For this Q1 release, only the spectra of objects with $\HE \le 22.5$ were extracted. At fainter magnitudes, the catalogues  derived from NISP imaging suffer from significant contamination by fake sources, due to the persistence effect in infrared detectors. Spurious residuals from preceding spectroscopic exposures are detected on the following photometric images, preventing the SIR PF from reliably extracting source positions at $\HE > 22.5$ (see \citealp{Q1-TP003} for a description of the NIR PF and the persistence phenomenon). Before proceeding with the redshift estimate, pixels that were flagged as `do not use' by the SIR PF are discarded (see \citealp{Q1-TP006} for details on flags). Following tests performed on simulated data and early spectra to eliminate as many artificial features as possible, only pixels that are built from at least three exposures are retained. This is the minimum number to guarantee a proper performance of the SIR PF sigma-clipping combination.

\begin{figure}[htbp!]
\centering
\includegraphics{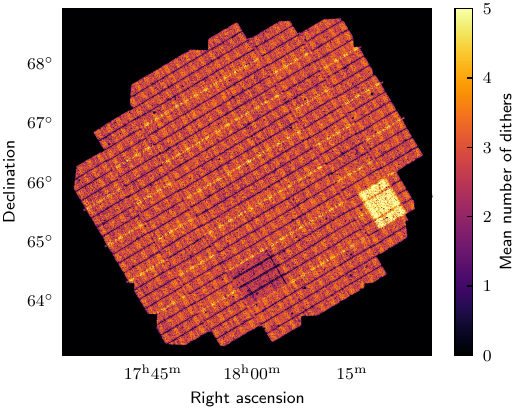}
\includegraphics{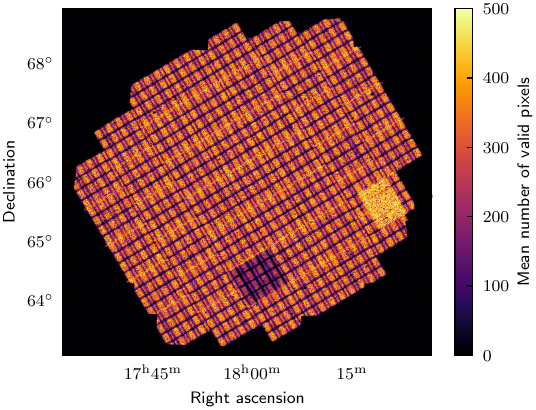}
\caption{\textit{Top}: Map of the mean number of dithers in the EDF-N field. \textit{Bottom}: Map of valid pixels.}
\label{fig:EDFNspectra}
\end{figure}

We also discard pixels that have the sixth bit of the quality flag activated: this corresponds to pixels at the edges of the spectrum, for which the absolute flux calibration curve has values close to zero, making calibration uncertain. Because of the current SIR configuration, this flag limits the available redshift interval to 0.9--1.8 in this release.
Accounting also for the various edge effects (size of the fields, focal plane geometry, and gaps between detectors -- see Fig.~\ref{fig:EDFNspectra}) yields a sample of 3.8 million processed spectra, among the 5.1 million objects 
in the $\HE\le 22.5$ catalogue, over the 63.1\,deg$^2$ Q1 area. This includes all spectra with at least one valid pixel.  To approach a science-grade sample, a lower threshold to the number of valid pixels is clearly required; however, at this stage, we chose not to apply any such cut, since future improvements in the pipeline may lead to different values for the optimal selection to be used. The nominal length of a spectrum that would not contain any invalid pixel is about 470 pixels.

It is important to note that this sample of processed (and delivered) spectra is in fact nearly two orders of magnitude larger than the number of $\ha$ emitters whose redshifts are expected to be measurable at the depth of the EWS, within the Q1 area.  Following the original science requirements of the mission, the NISP spectrograph was designed as to secure a redshift for an average number of $1700$ galaxies per deg$^{2}$, at a flux limit of $f(\ha)>2 \times 10^{-16}\,\ecgs$. Because of measuring redshifts using only the \ha\ line detection (see discussion in the following sections), by construction, these galaxies will have a redshift within the target range of $0.9<z<1.8$.  In practice, this implies that, within the Q1 release, we would expect at best to recover a total of $63 \times 1700 \simeq 100\,000$ redshifts within the target range, which is about 2\% of the total number of spectra extracted and delivered in Q1. The expected number of $\ha$ emitters, in fact, should be even smaller, if we consider that the requirements are based on extracting all sources with a broad-band magnitude of $\HE<24$, which is 1.5\,mag deeper than the current selection. Thus, any scientific investigation based on these data requires a strong and careful selection, as we shall discuss in the following sections.

 To allow this selection, the catalogue named \texttt{spectro\_zcatalog\_spe\_quality} contains all quality parameters characterising the spectra. These include the number of valid pixels (\texttt{spe\_npix}), the minimum and maximum wavelength (\texttt{spe\_w\_min} and \texttt{spe\_w\_max}), the maximum number of individual exposures used during the combination (\texttt{spe\_n\_dith\_max}), the median number of exposures calculated over the pixels of each spectrum (\texttt{spe\_n\_dith\_med}), and all the flags indicating processing errors during the redshift measurement for the three classes of the classification. In Fig.~\ref{fig:EDFNspectra}, we show how \texttt{spe\_n\_dith} and \texttt{spe\_npix} vary across the Euclid Deep Field North (EDF-N). The pattern resulting from the gaps between detectors is evident and this is only partly mitigated by the \Euclid dithering strategy. An area corresponding to a lower-quality observation is apparent at about \ang{18;00;} and \ang{+64;30;}, while an additional observation, executed to correct an uncovered area, shows an excess of number of dithers and valid pixels, at about \ang{18;20;} and \ang{+65;30;}. This clearly affects the ability to recover a redshift from spectra that fall, even in part, within these boundary areas.
 
The SPE PF employs a modified version of the Algorithm for Massive Automated Z Evaluation and Determination (\texttt{AMAZED}; \citealp{Amazed_2019}), tailored for \Euclid spectroscopic properties,  to calculate redshifts from spectra. This provides a broad classification of spectra into three categories: galaxies, stars, and quasars. For each category, dedicated models are fitted using a least-squares-fitting algorithm, yielding a probability distribution function (PDF) and a Bayesian evidence estimate for each category. The models for the galaxy category are a combination of six Bruzual--Charlot \citep[BC03;][]{BC03} continua, including interstellar absorption, and 14 emission-line-ratio templates built from the Virmos VLT Deep Survey (VVDS, \citealp{LeFevre2013}). The initial list of BC03 contained the 39 models described in \citet{Tremonti2003}. The 13 star formation histories included 10 instantaneous bursts at 5, 25, 100, 290, 640, 900, 1400, 2500, and 11\,000 Myr, one 6-Gyr-old continuous star formation episode, and two 12-Gyr-old models with exponentially declining star formation, the time constants being $\tau = 5\,{\rm Gyr}$ and $\tau = 9\,{\rm Gyr}$. For each of these models, three metallicity values were used: $0.2\, Z_\odot$, $Z_\odot$, and $2.5\, Z_\odot$. The first test made on \Euclid simulated spectra showed that, given the resolution and expected sensitivity, the metallicity diversity does not have a detectable impact on the redshift-measurement success rate. In addition, some instantaneous burst ages were very close to each other over the short wavelength range of \Euclid, so we finally built the used list by keeping the six templates that were used the most and after checking that this reduced list did not induce any modification in the results. This list includes five single-burst histories at 5, 25, 640, 1400, and 5000\,Myr and the exponentially declining history with $\tau=5\,{\rm Gyr}$. 

The 14 emission line ratio templates cover a large spread of the ratios measured between the main emission lines in the VVDS, as well as the possible presence of fainter emission lines detected only in specific galaxy spectral types. These templates  require compatible relative intensities in addition to positions of lines, and then allow fewer degrees of freedom for the fit. This prevents redshift solutions for which noise fluctuations of the spectra would be independently fitted by unrealistic emission lines. The quasar model is a single template built from the Sloan Spectroscopic Digital Survey (SDSS) mean quasar spectrum \citep{VandenBerk_2001}, while for stars the models are a collection of 36 stellar spectra from the European Southern Observatory 
Library of Stellar Spectra for spectrophotometric calibration \citep{Pickles_1998}. These templates include one template for O, A, and B stars, two for F stars, two G star templates, 15 for K stars, and 12 for M stars. The cold stars are more represented because of their diversity in the spectral domain of \Euclid. Ultra-cool dwarf templates (L and T) will be included in future releases. The final classification is made by selecting the category that shows the best evidence. This is collected in the 
\texttt{spectro\_zcatalog\_spe\_classification} catalogue, together with the probability associated with each category. It should be noted that, at this stage of development and test of the spectroscopic pipeline, the quality of the classification is still being improved (see Sect.~\ref{sc:classification}). 

For each category, the best-redshift solutions are identified by the most prominent peaks in the PDF. Up to five solutions are provided for each category and are listed  in three catalogues named \texttt{spectro\_zcatalog\_spe\_{\it TYPE}\_candidates}, where \texttt{\it TYPE} can be any of \texttt{\it galaxy}, \texttt{\it star}, or \texttt{\it qso}. Each object, identified by its \texttt{object\_id} as assigned by the MER PF pipeline, appears as many times as the number of redshift solutions, each identified by the \texttt{spe\_rank} parameter, ranging from 0 (best solution) to at most 4. These solutions are provided for all categories, independent of the actual classification result. All the templates (continuum and eventually emission line ratio) that correspond to each solution are also stored and available in the Science Archive. 

\begin{figure*}
\centering
\includegraphics[angle=0,width=1\hsize]{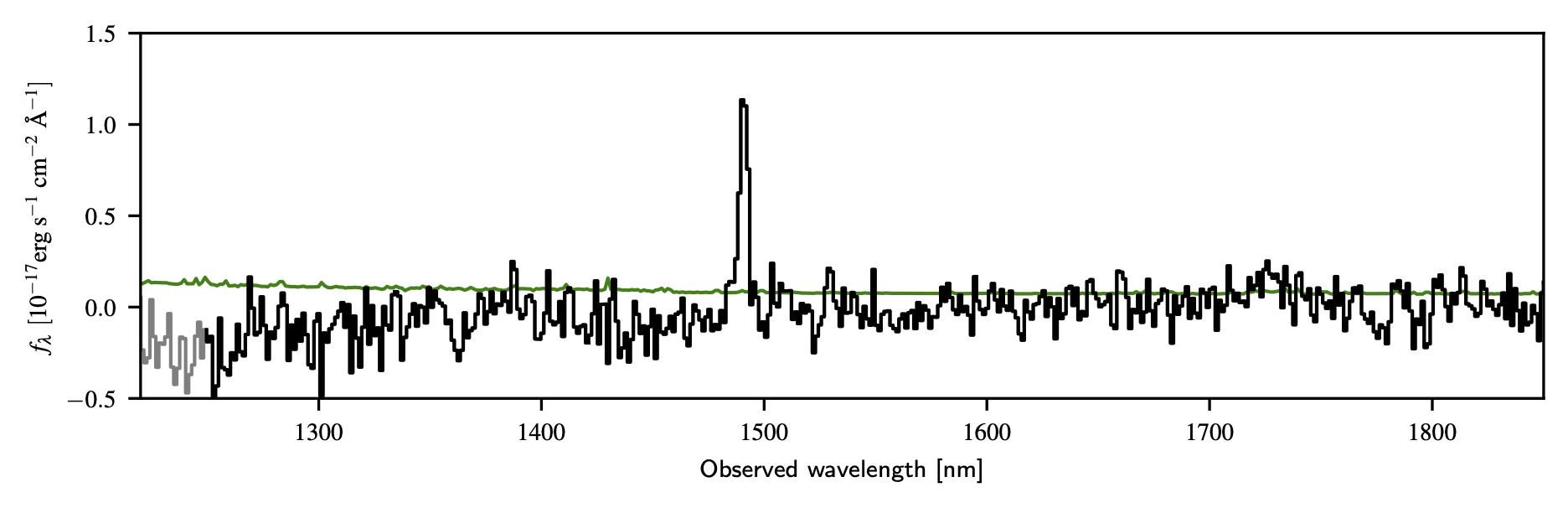}
\caption{\label{fig:spectrum} Typical spectrum (black line, uncertainties in green) of a galaxy from the EWS target sample, showing no continuum and a single emission line. From external spectroscopy, we know that this is \ha\ at $z=1.2711$; however, as is discussed in the text, SPE interprets the line as \oii\ in this case and assigns an erroneous redshift.}
\end{figure*}

\subsection{\label{sc:Haprior}The prior used for PDF computation}

The goal of the EWS survey is to compute clustering statistics for galaxies in the range $0.84<z<1.88$, using detections of the \ha\ line at $f(\ha)\,\ge 2\times10^{-16}\,\esc$ to secure their redshifts. In the present paper, however, we work with a slightly more conservative selection specific to Q1: $0.9 < z < 1.8$. This constraint will be relaxed in future  releases. The requirement, as is defined in the original design of the survey, is that, within this target range, the SPE PF will deliver a sample that is 45\% complete and 80\% pure, for objects with an \ha\ line detected with a signal-to-noise ratio (S/N) greater than 3.5, for sources with FWHM  $\leq 0.5$\,arcsec 
(since the spectral resolution depends on the size of the object along the dispersion direction). The SPE PF therefore has to supplement the measured redshifts with all the information on the measurement quality required to select a catalogue within the target range. This means supplying the correct redshift for at least 45\% of the full population of \ha\ emitters matching the selection criteria, as well as a percentage of less than 20\% of spurious objects with incorrect redshifts.
Given the wavelength coverage of NISP and the exposure time, \ha\ is the only line detected with good S/N for a large fraction of this redshift range and for a large fraction of the targeted galaxies. In addition, the continuum is barely detected, such that a galaxy at redshift $z=1.2711$ with $\HE = 21.5$ and  an \ha\ line with flux $5\times10^{-16}\,\esc$ has a typical spectrum as the one shown in Fig.~\ref{fig:spectrum}.

The problem is that, when using an agnostic line identification tool, the emission line in this case is not identified as \ha, but is incorrectly interpreted as \oii. In order to circumvent this issue, SPE PF introduces a prior in the calculation of the PDF, which favours solutions containing the \ha\ line. The prior acts on the PDF by dividing its value by a tunable factor for redshifts at which the \ha\ line can be detected. It currently has a value of 1000 below $z=0.9$ and above $z=1.8$, where \ha\ is not observed by the EWS, and unity for $0.9<z<1.8$, where it is. 

We thus ensure that, for objects such as the one shown in Fig.~\ref{fig:spectrum}, the redshift solution favoured by SPE is the correct one. Of course, sometimes this will induce the opposite effect: real \oii\ emission lines can be wrongly identified as \ha. However, to enter the red grism wavelength range, the \oii\ doublet at 3728\,\AA\ needs to come from a galaxy at $z>2.2$. At the limiting magnitude of the EWS, the number of such galaxies is lower than that of true \ha\ emitters within the target range by a factor of about 3.5 \citep[as estimated from the EL-COSMOS dataset;][]{Saito2020} and the net result is thus an improvement of the redshift measurement success rate.

\subsection{Reliability of the redshift estimation}
Associated with each solution, the \texttt{spe\_z\_prob} parameter gives the value of the integral of the PDF under the corresponding peak, over a $\pm\,3\,\sigma$ interval around the central value, where $\sigma$ is the width of the Gaussian fit to the peak. This provides an indicator of the reliability of the redshift estimate: a spectrum with two or more strong coherent emission lines will typically yield a PDF with a single peak whose integral is close to unity. Conversely, a spectrum such as the one in Fig.~\ref{fig:spectrum} will typically induce a PDF with multiple peaks, each corresponding to different potential identifications of the line. In this case, the integral of the PDF is shared among the peaks, leading to lower associated probabilities. Finally, for a spectrum without any prominent feature, the PDF will contain many faint peaks corresponding mostly to noise fluctuations and the probability associated with each solution will be low.  

The results for the success rate of the observed samples (described in Sect.~\ref{sc:Purity}) show that the \texttt{spe\_z\_prob} parameter is highly non-linear and that values below about 0.99 correspond to unreliable solutions. Improvements in the reliability of the redshift estimations are under development using machine-learning algorithms that will evaluate the whole shape of the PDF rather than a single peak. However, training of these algorithms requires a large sample of objects with secure redshifts, which will be provided by measurements on the EDS. An alternative will be to use simulated data, provided that they are sufficiently representative of the real spectra.

\section{\label{sc:perfs}Redshift estimate performance}

\subsection{Comparison with external redshift measurements}

\label{sc:comp}

In future \Euclid public releases, the completeness and purity of galaxy samples for cosmological investigations will be quantified using the Euclid Deep Survey (EDS) as a reference. This will provide us with an internal, fully self-consistent control sample, to quantify redshift measurement systematics that can affect galaxy clustering and the resulting inferred cosmological parameters. Lacking such an internal reference and given the broader nature of the Q1 dataset, to test the quality and limitations of the released data, we need here to resort to external measurements. The only appropriate dataset, in terms of area and redshift coverage, is provided by the Dark Energy Spectroscopic Instrument (DESI) Early Data Release catalogue \citep{Adame2024}, which partly covers the EDF-N.

 Within the DESI catalogue, we first selected objects with a reliable redshift, corresponding to: (1) quality flag \texttt{z\_warn}$\,=0$ (no issue identified in the redshift measurement); and (2) \texttt{Delta\_chi2} $>45$, which ensures that there is a significant difference between the main solution and the second one. The resulting sample was then matched to \Euclid objects featuring three or more spectroscopic dithers and a reliable detection in the MER catalogue, using a matching radius of \ang{;;0.2}. This produced a sample of 25\,469 matched objects. In the DESI catalogue, 10\,885 of these are identified as stars, 985 as quasars, and 13\,599 as galaxies.

\begin{figure}
\centering
\includegraphics{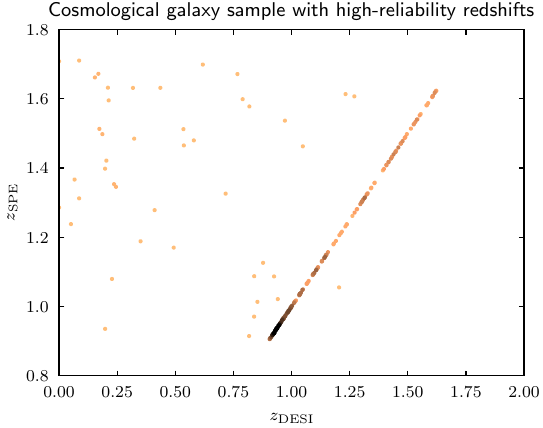}
\caption{\label{fig:z_vs_z_cosmo} Comparison between SPE and DESI redshifts for a selection corresponding to a baseline \Euclid cosmology sample of \ha\ emitters with SPE probability larger than 0.99. This plot is discussed extensively in Sect.~\ref{sc:purity_cosmo}.}
\end{figure}

\begin{figure}
\centering
\includegraphics{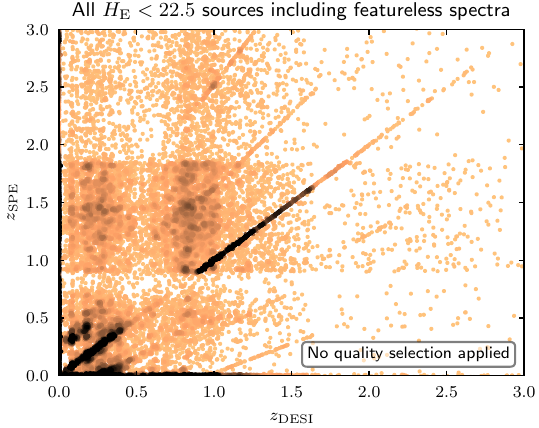}
\includegraphics{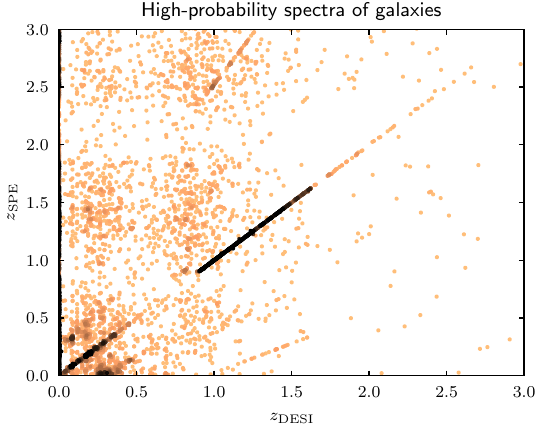}
\caption{\label{fig:z_vs_z} Comparison of SPE redshift measurements with those for all objects in common (including non-emitting galaxies and stars) from the DESI Early Data Release. \textit{Top}: All Q1 objects with at least three observed \Euclid dithers. This panel deliberately does not include any quality selection, so as to evidence the impact of the \ha\ prior (Sect.~\ref{sc:Haprior} and \ref{sc:comp}) and the limitations of spectroscopic data in the Q1 release beyond the \Euclid cosmology selection of Fig.~\ref{fig:z_vs_z_cosmo} (see Sect.~\ref{sc:z_failures}).
\textit{Bottom}: Same as top panel, but now selecting only high-quality objects classified as galaxies (i.e. with a SPE redshift probability $> 0.99$ and line width strictly narrower than the prior limit of $680\,\kms$). The colour coding indicates the local density of points, with black corresponding to the highest density. 
}
\end{figure}

Figures~\ref{fig:z_vs_z_cosmo} and \ref{fig:z_vs_z} 
compare the DESI and SPE redshift measurements, for various quality selections. Figure~\ref{fig:z_vs_z_cosmo} gives a first view of the quality of a \Euclid `cosmological sample' of \ha\-emitting galaxies, selected within the reference flux and redshift limits. This plot directly proves the current ability of SPE to recover redshifts for those galaxies that are the targets for which \Euclid NISP was built. A more in-depth discussion of this selection is presented in Sect.~\ref{sc:purity_cosmo}.

In the remainder of this section, we focus on the two panels of Fig.~\ref{fig:z_vs_z}, which deliberately look at the data for any kind of object, varying the quality thresholds.  The scope here is to test the ability of \Euclid and the SPE pipeline in particular to work out of their `comfort zone' and recover information beyond the cosmological goals NISP was designed for. The matched DESI sample in fact also includes galaxies outside the \ha\ visibility range, for which the spectral characterisation has to be based on emission lines other than \ha\ or on the continuum;  there are also stars, for which some information can also be present in terms of absorption lines.

In the upper panel of Fig.~\ref{fig:z_vs_z}, the effects of the previously discussed  \ha\ prior are easily noticeable as an over-density of points in the $0.9 < z_{\rm SPE} < 1.8$  region.  
Here, no quality criteria are applied, in order to give a full illustration of the various artefacts that can affect the performance of the SPE PF, in particular when dealing with measurements outside the target redshift range.

The bottom panel displays the same information, but only for galaxies with \texttt{spe\_z\_proba}$\,\ge 0.99$, showing the relative efficiency of this selection in excluding incorrect solutions. It must be noted as well that, even if the percentage of correct solutions is low in the $z\,=\,$0\,--\,0.9 interval (see Sect.~\ref{sc:Purity} for detailed numbers), about 15\% of galaxies measured by DESI in this redshift interval have a redshift measured by SPE in the same interval with \texttt{spe\_z\_proba}$\,\ge 0.99$. Considering that the redshifts measured by SPE cannot rely on any strong emission lines except for specific categories (such as active galactic nuclei (AGNs)), these measurements are solely based on a small portion of the continuum. This is nevertheless interesting, since NISP and the EWS were designed to measure redshifts for \ha\ emitting galaxies at $z>0.9$, and limited performance is expected below this redshift.  

\subsection{Redshift measurement uncertainties}

\begin{figure}
\centering
\includegraphics{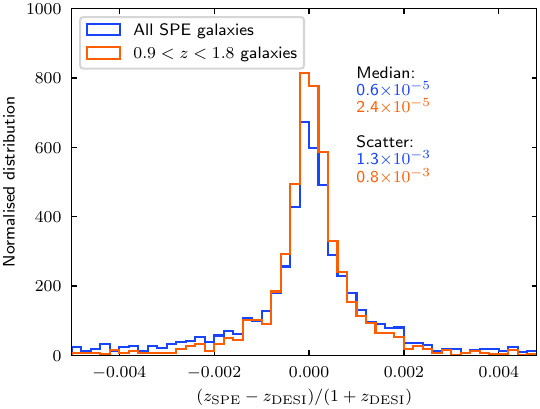}
\caption{\label{fig:z_precision} Normalised distribution of $(z_{\rm SPE} - z_{\rm DESI}) / (1+z_{\rm DESI})$ for all the objects classified by SPE as galaxies (blue) and only in the target redshift range ($0.9<z<1.8$, orange).}
\end{figure}

To evaluate the precision of \Euclid redshifts, we computed the relative redshift difference $(z_\mathrm{SPE} - z_\mathrm{DESI}) / (1+z_\mathrm{DESI})$ between the redshift measured by SPE, for objects classified as galaxies and the one provided by DESI. After excluding outliers, defined as those having an absolute difference larger than 0.005, we obtain a distribution with a median value of $0.6\times10^{-5}$ and a $1\,\sigma$ dispersion of $1.3\times10^{-3}$, which is plotted in Fig.~\ref{fig:z_precision}. This provides an upper limit on the \Euclid redshift uncertainties, since this quantity also contains the DESI uncertainties. However, with the DESI resolution ranging between 2000 and 5500, that is 4--5 times better than NISP, \Euclid errors should dominate this comparison. If we restrict the sample to galaxies within the \Euclid target range ($0.9<z<1.8$, but without flux restriction), to ensure that at least the  \ha\ emission line is used for the redshift determination if it is present, the RMS error drops to $7\times10^{-4}$, with a median value of $2.4\times10^{-5}$.  If we additionally cut the sample at  the nominal  \ha\ flux limit of $2\times10^{-16}\,\esc$, the dispersion further reduces to $4.9\times10^{-4}$. Given the low resolution of NISP and the sizeable spectroscopic undersampling, this remarkable agreement confirms the quality of the wavelength calibration, as is described in \cite{Q1-TP006}.

\subsection{Types of redshift failure}

\label{sc:z_failures}

As is shown in Fig.~\ref{fig:z_vs_z}, a significant fraction of galaxies have an incorrect redshift if only basic quality criteria are applied, and are thus not located on the one-to-one relation. There are two types of incorrect redshift visible in this figure. 

A first type, which in the forthcoming \cite{Risso2025} paper are called `line interlopers', is caused by the misidentification of a detected line. As was already explained in Sect.~\ref{sc:Haprior}, the SPE pipeline can attribute the wrong line to a feature. This is particularly the case if only one line can be seen in the \Euclid spectral range. In this situation, the relation between the true redshift, $z_{\rm true}$, and wrong redshift, $z_{\rm misID}$, is
\begin{equation}
\frac{1+z_{\rm true}}{1+z_{\rm misID}} = \frac{\lambda_{\rm misID}}{\lambda_{\rm true}},
\end{equation}
where $\lambda_{\rm true}$ and $\lambda_{\rm misID}$ are the true and wrong (misidentified) rest-frame wavelengths of the measured line, respectively. In Fig.~\ref{fig:z_vs_z}, we see the signature of a few possible line misidentifications as the linear tracks off the main diagonal of the plot. Clearly, if all redshifts in the Q1 release are used without any quality filter (bottom panel of Fig.~\ref{fig:z_vs_z}), a large variety of misidentifications are possible. 

If, as in the bottom panel, we select high-quality SPE galaxies (defined as having \texttt{spe\_z\_proba}\,$\ge\,0.99$ and a line width -- expressed in terms of velocity dispersion -- smaller than 680\,km\,s$^{-1}$ to exclude spurious broad features), we are left with three such features indicating line misidentifications. The one above the one-to-one diagonal corresponds to $(1+z_{\rm SPE})/(1+z_{\rm DESI}) \simeq  1.76$: this results from cases in which SPE has interpreted a line as \oii\ when in fact it is \ha. We can also notice two discrete features below the one-to-one relation, corresponding to \ha\ being misidentified as \Pa, $(1+z_{\rm SPE})/(1+z_{\rm DESI}) \simeq 0.51$, or \siii{9530}, $(1+z_{\rm SPE})/(1+z_{\rm DESI}) \simeq 0.68$. Finally, inspection of the $(1+z_{\rm SPE})/(1+z_{\rm DESI})$ histogram reveals the presence of a small peak at 1.34, corresponding to \ha\ being misidentified as \hb, although we surprisingly do not find the peak corresponding to \oiii{}. 

As can be seen by comparing with Fig.~\ref{fig:z_vs_z_cosmo}, such misidentifications are largely suppressed by the \ha\ prior, and this aspect will be revisited in future versions of the pipeline. We remark that no obvious misidentification is observed in the target redshift range for cosmology, $z=0.9$--1.8.

The second type of error (giving rise to what we call `noise interlopers' in \citealp{Risso2025}) is caused by noise features or pipeline artefacts, such as unidentified residuals of the zeroth order signal, which are mistaken for real lines. In such cases, the measured redshift values are spread over a broad range. However, they are less likely to happen for redshift ranges where two real lines fall within the nominal NISP wavelength  range, since this would require two spurious peaks separated by the same exact wavelength difference of two emission lines (this is the case in the range $1.4<z<1.8$ when considering \oiii{5007} and \ha). Conversely, the \ha\ prior tends to push these spurious redshifts into the cosmology target range. As a result, the distribution of these incorrect redshifts is not uniform, as is shown in Fig.~\ref{fig:z_vs_z}.

\section{\label{sc:classification} Spectral classification}
\subsection{Broad categories}
In addition to the redshift, SPE also provides classifications into three main types: galaxies, stars, and quasars. We compared these classifications in the EDF-N with the DESI ones. In Table~\ref{tab:conf_matrix}, we present the confusion matrix of the SPE classifier, which describes how the objects from a given `true' class are classified into the various classes by the classifier. For a perfect classifier, we would have 100\% on the diagonal and 0\% everywhere else. We assume here that DESI is the ground truth, which could lead to biases in the estimate of the confusion matrix.

We consider two different cases. For the first, we selected high-quality spectra with three or more dithers on average (see Sect.~\ref{sc:basic_description}). For the second case, we additionally required a class probability higher than 0.99. The results are not significantly better for the second case. This suggests that the spectral quality and the SPE classification probability criteria alone cannot yet efficiently select objects with a good classification.

The results are satisfactory for galaxies, with around 80\% success. The results for quasars are poor, with about 80\% of them being classified as galaxies. The current version of the pipeline uses a single template for quasars based on the average spectrum of SDSS quasars. It thus allows too little flexibility to fit the \Euclid spectra reliably. In addition, this template has broad lines, and the observed type~II AGNs tend to be fitted better by the galaxy templates, for example when the line-spread function is even slightly overestimated.

Only about 50\% of the stars are correctly classified, while the remaining 50\% leak into the galaxy class. Only 23\% of these false galaxies are at $z<0.1$, and 21\% of them are in the target redshift range for cosmology ($0.9<z_{\rm SPE}<1.8$). This emphasises the fact that additional non-SPE criteria will be necessary to minimise the contamination of the cosmological sample by stars and the high-quality imaging of the \Euclid space telescope will undoubtedly be efficient in doing this. In contrast, about 75\% of the objects classified as stars by SPE are DESI stars. The selection of star samples is thus very promising and the combination with other quality criteria from the \Euclid photometric pipeline should provide high-quality samples.

\begin{table}
\centering
\caption{\label{tab:conf_matrix} Confusion matrices of the OU-SPE classification.}
\begin{tabular}{lrrr}
\hline \hline
\noalign{\vskip 2pt}
Classified by SPE as: & galaxies & stars & quasars \\
\hline
\noalign{\vskip 2pt}
\multicolumn{4}{c}{High-quality \Euclid spectra}\\
\multicolumn{4}{c}{($\ge$3 dithers in average)}\\
\hline
\noalign{\vskip 2pt}
DESI galaxies & 80\% & 12\% & 8\%\\
DESI stars & 49\% & 49\% & 2\% \\
DESI quasars & 77\% & 7\% & 16\%\\
\hline
\noalign{\vskip 2pt}
\multicolumn{4}{c}{High-quality \Euclid spectra and}\\
\multicolumn{4}{c}{SPE classification probability $>$0.99}\\
\hline
\noalign{\vskip 2pt}
DESI galaxies & 84\% & 10\% & 6\%\\
DESI stars & 49\% & 50\% & 1\% \\
DESI quasars & 79\% & 7\% & 14\%\\
\hline
\end{tabular}
\end{table}

\subsection{Galaxy sub-categories}
In the \texttt{spectro\_zcatalog\_spe\_galaxy\_candidates} catalogue, the SPE PF provides a \texttt{spe\_subclass} parameter, which is a short name for the emission line ratio templates used for the redshift calculation. The names refer either to templates based on morphological types (`Irr', `Scd', `Sbc') eventually coupled with additional properties derived from VVDS spectra such as `SB' for starburst or `BX' for excess in the blue. The most represented templates on the full sample or on the sample selected by  $\texttt{spe\_z\_proba}\ge 0.99$ are either of starburst or `BX' types. However, given the short wavelength range of NISP, the bias towards the solutions given by the \ha\ line and the fact that only fluxes greater than $2\times10^{-16}\,\esc$ are detectable, this distribution is surely not representative of the reality of the observed population. 

As for the continuum templates, the ones corresponding to recent single star formation episodes (5, 25, and  640 Myr) represent more than half of the sample. The fraction rises to 75\% for the object with a correct redshift as compared to DESI. This confirms that the way the redshift is measured favours galaxies with high star formation rates.

\section{\label{sc:LineFlux}Line-flux measurement products and performance}
\begin{figure*}
\centering
\includegraphics[width=18cm]{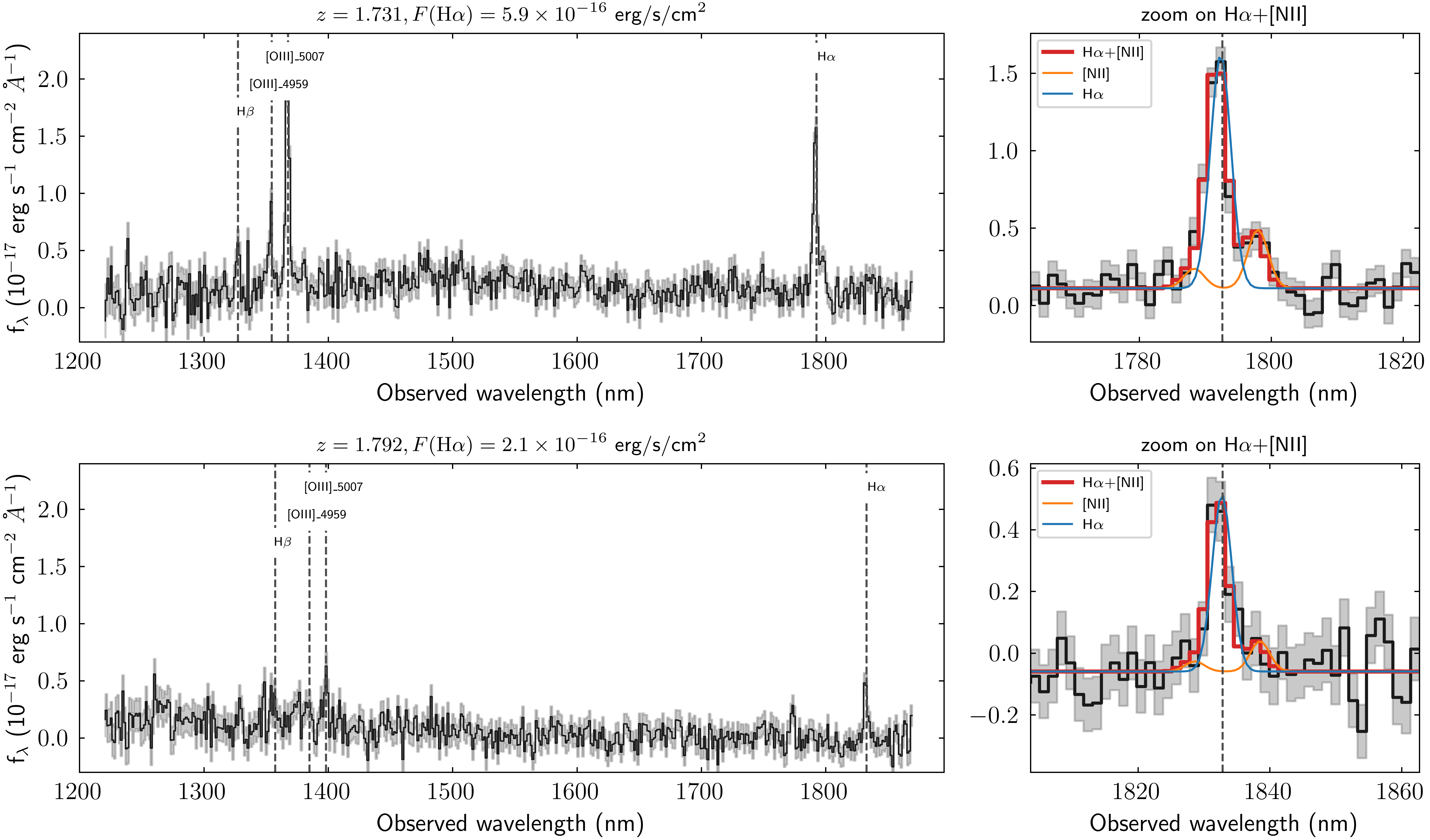}
\caption{\label{fig:LineFit} Illustrative examples of the line flux measurement performed in SPE. Two spectra are shown from the EWS at $z\simeq1.7$--1.8, with fluxes of $f(\ha)\simeq2\times10^{-16}$ and $5.9\times10^{-16}\,\esc$ (\emph{upper} and \emph{lower} panels, respectively). The \emph{left} panels show the entire spectrum, highlighting the position of the main emission lines, and the \emph{right} ones the results of the multiple Gaussian fit to the \ha+\nii\ complex showing in different colours the contribution of the \ha\ line (in blue), of the \nii\ doublet (in orange), and their combination (in red).}
\end{figure*}

Once a redshift solution has been estimated for the galaxy model, the fluxes of all emission and absorption lines that are expected to be present in the spectrum are measured. It must be noted that for this Q1 release, even if an object is classified as a quasar or a star, the lines are identified at the redshift of the galaxy solution and therefore should not be trusted; this will be modified in future releases. Line fluxes are measured with two methods: `direct integration' provides fluxes integrating the values of all the pixels of the emission line above the continuum; and `Gaussian fit' derives them from the fit of the line using a single, double, or multiple Gaussian model, depending on the line considered.  Given the low spectral resolution of NISP, the \ha\ and \nii\ doublet lines direct-integration fluxes are measured jointly over a unique wavelength interval to produce a single flux measurement, while the Gaussian fit provides fluxes separately for the \ha\ and \nii\ doublet. 
Examples of the fitting procedure results are displayed in Fig.~\ref{fig:LineFit} for a high-S/N emission line, as well as for one close to the flux detection limit.

For each line and each method, SPE provides the flux, the S/N, the central wavelength, and the full width at half maximum. All measurements relative to emission lines are stored in the {\small\texttt{spectro\_line\_features\_catalog\_spe\_line\_features\_cat}} table: for each object, this contains all the detected lines for each redshift solution. A given line can be identified either by its name (e.g. \texttt{Halpha}) or a numerical identifier, and the link between line names and identifiers is given in the \texttt{spectro\_model\_catalog\_spe\_lines\_catalog} table. Therefore, to retrieve the flux of the \ha\ line for the main redshift solution of all objects, the selection should include \texttt{spe\_line\_name}$=$`Halpha' and \texttt{spe\_rank}=0, together with any additional conditions. 

A S/N value of 0.0 indicates that the measurement is unreliable, either due to unrealistically small value of the measured flux or to the fact that the flux measurement is made over a single valid pixel. Excluding those doubtful measurements, there are 272\,676 objects for which the \ha\ line is formally detected and has a flux measured with the Gaussian-fit method greater than $5\times10^{-17}\,\esc$. This value is the limit under which no measurement can be performed, the limit for a $3\sigma$ detection being much higher:  in the interval $[2\times 10^{-16}$, $2.5\times 10^{-16}]\,\esc$, the average S/N is 6, with a dispersion of $2.7$, meaning that more than 70\% of these lines are detected with $\textrm{S/N}>3.5$. This value is obtained with all lines identified as \ha\ whatever the quality of the measurement. For objects with \texttt{z\_spe\_proba}$\,\ge 0.99$, the average value of the S/N rises to 6.7.

\section{\label{sc:Purity} Success rates}

\begin{figure}
\centering
\includegraphics{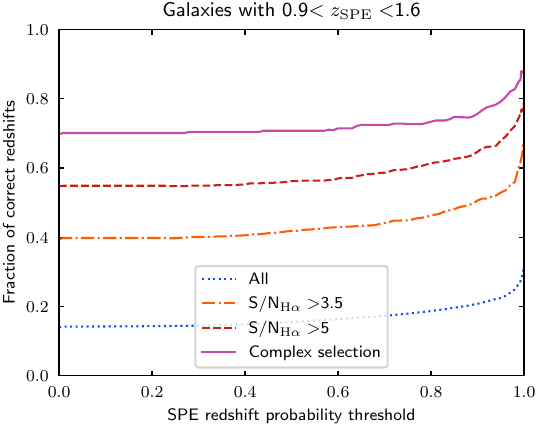}
\caption{\label{fig:purity} Success rate (fraction of correct redshifts) as a function of the SPE redshift probability threshold for objects classified as galaxies and with a measured redshift between 0.9 and 1.6. The various solid coloured lines correspond to all objects (dotted blue line), objects with an \ha\ S/N above 3.5 (dash-dotted orange line), and objects with an \ha\ S/N above 5 (dashed red line). Finally, the solid purple line presents the result of the following selection: \ha\  S/N$\,>\,$5, measured \ha\ flux larger than $2\times10^{-16}\,\esc$, and a line width strictly smaller than the upper bound of the prior ($680\,\kms$).}
\end{figure}

Using the DESI redshifts in the EDF-N as a reference, we can estimate the success rate of SPE redshift measurements. In this paper, we call the `success rate' the fraction of objects with a correct redshift for a given selection based on the measured parameters and the quality metrics returned by SPE.\footnote{ We could have called it `purity', but the notion of purity have a very strict definition in the \Euclid context and we thus choose another terminology to avoid any ambiguity.} Since the DESI redshift distribution is biased towards lower redshifts and brighter objects than the objects targeted by \Euclid, the results must be taken with caution, as discussed in Sect.~\ref{sc:comp}. We discuss the impact of the choice of DESI as reference sample in Appendix\,\ref{sect:DESI_impact}. A better internal assessment will be possible at DR1, when a subset of the EDS \citep{Scaramella-EP1} will be available.  The full EDS will benefit from 15 multiple EWS-like red grism (1206--1892\,\AA) visits, supplemented by a further 25 visits with the blue grism (900--1300\,\AA). This has been designed to retrieve virtually pure and complete reference samples at the EWS depth (thanks to larger sensitivity), minimal spectral confusion between adjacent objects (thanks to multiple observing angles), and reduced redshift degeneracies (thanks to the extended blue-plus-red wavelength range).
We will therefore have a reliable and complete control sample with which to calibrate EWS observations -- including measurements of \ha\ fluxes, which will allow for a proper estimate of the completeness. In addition, realistic simulations will also be released along with DR1.

To obtain a quantitative estimate of the success rate, relative to the DESI sample, we computed the fraction of objects with $|z_{\rm SPE} - z_{\rm DESI}| / (1+z_{\rm DESI}) < 0.003$ in various sub-samples. This range corresponds to the nominal $3\,\sigma$ interval around SPE measurements.  We focused on the range $0.9<z_{\rm SPE}<1.6$, which is the usual cosmological target range of \Euclid, where \ha\ falls within the red grism, but limited to $z=1.6$, corresponding to the upper cut-off of the DESI sample. 
In the following sections, we study the dependence of the success rate as a function of the SPE redshift-probability threshold (\texttt{gal\_spe\_z\_prob}). As was expected, a high threshold produces a more reliable sample. However, this reduces the size of the recovered sample and thus the completeness, which cannot be estimated yet in the absence of full-depth EDFs, and the cut-off in magnitude. 

\subsection{Success rate for non-cosmological science}

\label{sc:purity_legacy}

The success rate for various quality selections is shown in Fig.~\ref{fig:purity}. In the absence of any additional selection (blue), the reliability goes from 14\% for a null SPE redshift probability threshold to 27\% for a 0.99 threshold. The results can be improved dramatically by adding \ha\  S/N criteria (3.5 in orange and 5 in red). With these additional criteria, we reach 61\% and 75\% success rates for a 0.99 probability threshold and S/N thresholds of 3.5 and 5, respectively. Finally, we add additional criteria on the flux and the width of the \ha\ line found by template fitting. We keep only the sources with \ha\ flux larger than $2\times 10^{-16}\,\esc$ and also discard all objects for which the estimated line width saturates at the maximum value of the allowed range ($680\,\kms$), which is a signal of a spurious broad feature mistaken as emission line. With this selection, we reach a success rate of 85\% for a $\texttt{gal\_spe\_z\_prob}=0.99$ threshold.

We also studied the SPE success rate outside the redshift range in which \ha\ can be detected. In Table~\ref{tab:perf_zslices}, we summarise the performance obtained in various redshift slices. If we consider all three classes of objects and use a 0.99 redshift probability threshold for the best class, we reach a success rate of 53\% at low $z$ ($z<0.1$) without adding any additional criteria. This result is encouraging, but it falls to only 7\% if we select objects classified as galaxies and we see that the apparently good results are mainly driven by stars (see Sect.~\ref{sc:classification}). Not surprisingly, the results are also poor at $0.1<z<0.9$, with an 8\% success rate, explained mainly by the absence of any strong spectral features in the red-grism wavelength range. The performance is even poorer above $z=1.8$, with a 2\% success rate. This result is not so surprising, since no bright spectral feature is expected to be detected in the red grism above $z=1.8$ given the sensitivity of NISP.

Our results illustrate the importance of applying appropriate selection criteria when using spectroscopic data from the Q1 release. Measurements for galaxies in the \Euclid target redshift range, where \ha\ is covered by the red grism provide encouraging results. The same is true for stars. Outside of this redshift range, SPE results alone cannot be used to build a reasonably pure redshift sample for statistical studies; they must be complemented with additional data such as photometry \citep{Cagliari24} or morphological criteria that will be provided by the high-quality \Euclid images. A visual inspection of the spectra can also help to refine small samples for specific studies (Euclid collaboration: Quai et al. in prep.).

\begin{table}
\centering
\caption
{\label{tab:perf_zslices} Summary of SPE success rate 
}
\begin{tabular}{lccc}
\hline
\hline
\noalign{\vskip 2pt}
Measured redshift & \multicolumn{3}{c}{Success rate}\\
\hline
\noalign{\vskip 2pt}
& $P_{\rm SPE}>$0.9 & $P_{\rm SPE}>$0.99 & $P_{\rm SPE}>$0.999 \\
\hline
\noalign{\vskip 2pt}
\multicolumn{4}{c}{All the objects}\\
\hline
\noalign{\vskip 2pt}
$z<0.1$ & 47\% & 53\% & 58\% \\
$0.1<z<0.9$ & \phantom{0}8\% & \phantom{0}8\% & \phantom{0}8\% \\
$0.9<z<1.6$ & 27\% & 34\% & 38\% \\
$0.9<z<1.8$ & 24\% & 31\% & 35\% \\
$z>1.8$ & \phantom{0}2\% & \phantom{0}2\% & \phantom{0}3\% \\
\hline
\noalign{\vskip 2pt}
\multicolumn{4}{c}{Objects classified as galaxies only}\\
\hline
\noalign{\vskip 2pt}
$z<0.1$  & \phantom{0}6\% & \phantom{0}7\% & \phantom{0}8\%\% \\
$0.1<z<0.9$ & \phantom{0}8\% & \phantom{0}8\% & \phantom{0}8\% \\
$0.9<z<1.6$ & 26\% & 33\% & 37\% \\
$0.9<z<1.8$ & 24\% & 30\% & 35\% \\
$z>1.8$ & \phantom{0}2\% & \phantom{0}2\% & \phantom{0}2\%\\
\hline
\noalign{\vskip 2pt}
\multicolumn{4}{c}{Cosmological galaxy sample (SPE classification)}\\
\hline
\noalign{\vskip 2pt}
$0.9<z<1.6$ &  64\% &  75\% &  82\% \\
$0.9<z<1.8$ &  59\% &  71\% &  78\% \\
\hline
\noalign{\vskip 2pt}
\multicolumn{4}{c}{Cosmological sample (galaxies only in input)}\\
\hline
\noalign{\vskip 2pt}
$0.9<z<1.6$ & 72\% &  83\% &  89\% \\
$0.9<z<1.8$ & 67\% &  78\% &  85\% \\
\hline
\end{tabular}
\tablefoot{The success rate is given for various measured redshift ranges (lines) and SPE redshift probability thresholds, $P_{\rm SPE}$ (columns). These rates are derived from a comparison with DESI data, and some care is required when interpreting them. DESI cannot detect emission-line galaxies beyond $z=1.6$, and therefore all our correct redshifts between 1.6 and 1.8 will not be included here. }
\end{table}

\subsection{Success rate for cosmology goals}

\label{sc:purity_cosmo}

\begin{figure}
\centering
\includegraphics{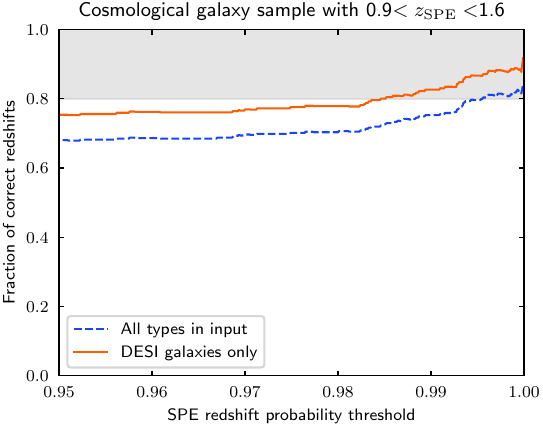}
\caption{\label{fig:perf_cosmo} Success rate (fraction of correct redshifts) as a function of the redshift probability for the cosmological sample. The dashed blue line is computed using all types of objects as input and SPE is used to identify the galaxies. The solid orange line is computed using objects classified by galaxies by both SPE and DESI. The grey area represents the \Euclid performance goal for cosmology.}
\end{figure}

{
The measurement of reliable redshifts between $z=0.9$ and $z=1.8$ is the first crucial step towards \Euclid's cosmological goals based on galaxy clustering measurements. Although we do not have a true reference sample to test the cosmological performance with, we can nevertheless draw useful indications by properly trimming the Q1 and DESI samples. 
We thus selected a galaxy sample similar to the baseline cosmological sample: \ha\ flux larger than $2\times10^{-16}\,\esc$ and $\textrm{S/N}>3.5$. In addition, we kept only objects for which the line width is narrower than the prior limit of $680\,\kms$ to exclude artefacts (Sect.~\ref{sc:perfs}). The size limit of \ang{;;0.5} is not included in this study and even better results could be obtained by adding this criterion.
 
In Fig.~\ref{fig:z_vs_z_cosmo}, we compare the redshift obtained by SPE and by DESI for the selection described in the previous paragraph and a SPE redshift probability larger than 0.99. Since the \Euclid selection for cosmology is based on the measured \ha\ line, the measured redshifts are all between 0.9 and 1.8. The results are dramatically improved compared to simpler selections discussed in Sect.~\ref{sc:perfs} with most of the galaxies on the 1:1 relation. However, we also find a small population of galaxies with a lower DESI redshift, while only one object has a higher DESI redshift. Since they are not aligned, we can reasonably assume that these are spurious SPE redshifts that do not arise from simple line misidentifications. These $z<0.9$ sources erroneously measured by SPE at $z>0.9$ presumably arise from noise in the \Euclid red-grism spectral range and it is not surprising to have a small fraction of such spurious redshifts out of the large population of galaxies with $\HE <22.5$ and $z<0.9$  (see Sect.~\ref{sc:basic_description}).

In Fig.~\ref{fig:perf_cosmo}, we present the SPE success rate as a function of the redshift properties for galaxies with measured properties matching the criteria of the cosmological sample. As is explained in Sect.~\ref{sc:purity_legacy}, we consider only the $0.9<z<1.6$ range, where both DESI and SPE are reliable. If we select galaxies based on the SPE classification only, we reach an 80\,\% success rate for \texttt{z\_spe\_proba}$\,\ge 0.996$ (dashed blue line). However, the SPE-only classification is not highly reliable (Sect.~\ref{sc:classification}) and we will eventually benefit from \Euclid morphological as well as photometric information to discard stars and lower-redshift galaxies from the sample. To obtain a performance forecast closer to the expected methods that will be used in \Euclid, we computed the success rate for a clean sample classified by both SPE and DESI as galaxies. A \texttt{z\_spe\_proba} threshold of 0.985 is enough to obtain an 80\% success rate and we can reach 89\% for a stringent \texttt{z\_spe\_proba}$\ge 0.999$ criterion.

In short, we have demonstrated that success rates higher than the specifications (see Table~\ref{tab:perf_zslices} for a summary) can be obtained for the cosmological sample if we select the sample carefully. For the future \Euclid DR1, further adjustments will be possible in order to establish the optimal threshold that yields the best compromise between success rate and sample size, in order to maximise the performance for cosmology. These future analyses will use the EDS data, which are not part of Q1.
}

\section{\label{sc:conclusion} Conclusions}

We have presented the first data release (Q1) of the SPE PF results. We have shown that, if we properly combine the various quality indicators provided by the pipeline, we can reach a success rate above 80\% in the redshift range targeted for cosmology ($0.9<z<1.8$). The redshift precision (approximately $10^{-3}$) and accuracy (better than $3\times 10^{-5}$) are also excellent. This is encouraging for the future DR1 and the first cosmological results, but the completeness will still need to be assessed when EDS data will be available and all EWS sources down to $\HE=24$ instead of 22.5 will have been processed.

As would be expected, the results are less good outside of this redshift range, since no strong spectral features are then observable within the NISP red-grism spectral range and these data should be used with caution for non-cosmology science. Future deep-field observations including the blue grism should dramatically improve \Euclid's performance in this regime. In addition, the sample of star candidates selected by SPE PF based on spectroscopy alone is 75\% pure. This is very promising for the study of cool stars \citep{Banados}.

Finally, the current results are only a first step and we can expect major improvements in the future. The SIR PF products injected into the SPE PF still contain important artefacts \citep{Q1-TP006}, which will be corrected in the next versions of the pipeline. This should dramatically reduce the fraction of spurious redshifts and enable us to use less stringent selection criteria, increasing the size of the high-reliability samples. The SPE PF pipeline itself will also improve. Deep-learning approaches have been developed to estimate the redshift reliability and classify the spectra. We already trained as well as tested such algorithms on simulations and they outperformed the current algorithm. However, we shall need real future EDS data to build a proper training set for the analysis of real EWS data. Finally, the selection of galaxies for cosmological analyses will also benefit from non-spectroscopic information (e.g. morphology or photometry), allowing us to obtain sufficiently pure and complete samples.

\begin{acknowledgements}
\AckEC 

This research used data obtained with the Dark Energy Spectroscopic Instrument (DESI). DESI construction and operations is managed by the Lawrence Berkeley National Laboratory. This material is based upon work supported by the U.S. Department of Energy, Office of Science, Office of High-Energy Physics, under Contract No. DE–AC02–05CH11231, and by the National Energy Research Scientific Computing Center, a DOE Office of Science User Facility under the same contract. Additional support for DESI was provided by the U.S. National Science Foundation (NSF), Division of Astronomical Sciences under Contract No. AST-0950945 to the NSF’s National Optical-Infrared Astronomy Research Laboratory; the Science and Technology Facilities Council of the United Kingdom; the Gordon and Betty Moore Foundation; the Heising-Simons Foundation; the French Alternative Energies and Atomic Energy Commission (CEA); the National Council of Science and Technology of Mexico (CONACYT); the Ministry of Science and Innovation of Spain (MICINN), and by the DESI Member Institutions: www.desi.lbl.gov/collaborating-institutions. The DESI collaboration is honoured to be permitted to conduct scientific research on Iolkam Du’ag (Kitt Peak), a mountain with particular significance to the Tohono O’odham Nation. Any opinions, findings, and conclusions or recommendations expressed in this material are those of the author(s) and do not necessarily reflect the views of the U.S. National Science Foundation, the U.S. Department of Energy, or any of the listed funding agencies.

\end{acknowledgements}

%
%
\bibliographystyle{aa}
\bibliography{Q1_TP007b}

\begin{appendix}

\section{Representativeness of the DESI reference sample}

\label{sect:DESI_impact}

\begin{figure}
\centering
\includegraphics[width=6.6cm]{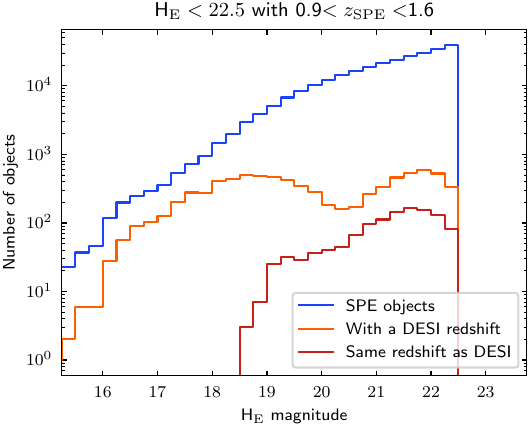} 
\includegraphics[width=6.6cm]{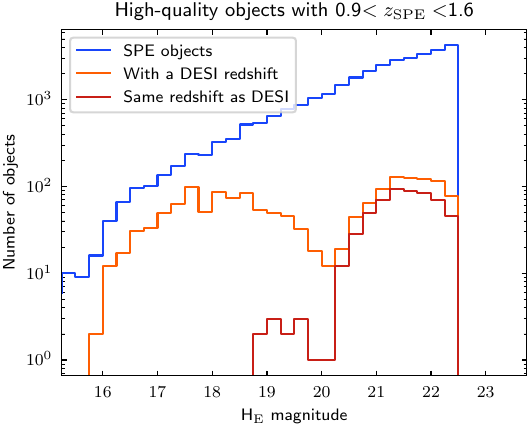} 
\includegraphics[width=6.6cm]{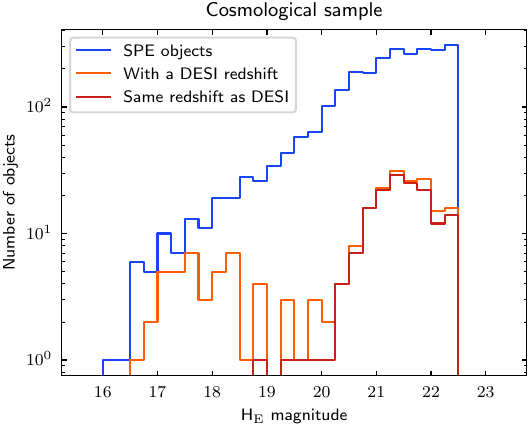}   \includegraphics[width=6.6cm]{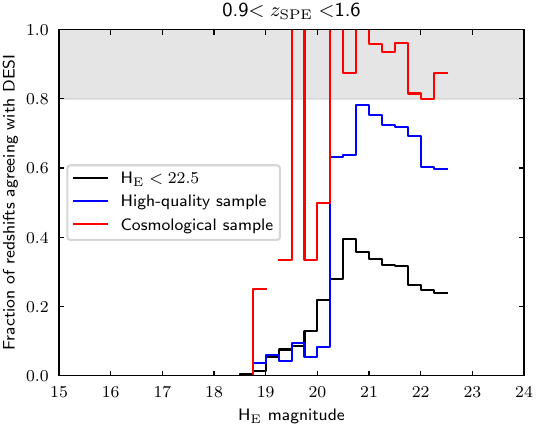} 
\caption{\label{fig:representativeness} Figures illustrating the representativeness of DESI as a reference sample. \textit{From the top :} Number of $0.9<z_{\rm SPE}<1.6$ objects as a function of the \HE magnitude, for the high-quality objects described in Sect.\,\ref{sc:z_failures} and for the cosmological sample described in Sect.\,\ref{sc:purity_cosmo} respectively. The blue, orange, and red histograms are the SPE sample, the sub-sample with a DESI redshift, and the sub-sample with a redshift agreeing with DESI, respectively.  \textit{Bottom:} Fraction of redshifts agreeing with DESI as a function of the \HE magnitude for the $0.9<z_{\rm SPE}<1.6$ SPE sample (black), the sub-sample with a DESI redshift (blue), and the cosmological sample (red).}
\end{figure}

 In Fig.\,\ref{fig:representativeness}, we present the \HE-magnitude distribution of various selections. If we consider the full SPE sample with \HE$<$22.5 and $0.9<z_{\rm SPE}<1.6$ (upper left panel), we can observe that number of SPE objects (blue) per magnitude bin increases steeply with increasing magnitude, while the number of objects with both a SPE and a DESI redshift (orange) is significantly flatter than for the full SPE sample above $\HE=20$. Below this magnitude, we can notice a bump peaking around $\HE=18$, as discussed below. Only a small fraction of the \HE-faint \Euclid galaxies can thus be compared with DESI. This confirms that DESI is not suited to estimate the completeness of the SPE sample. We find a similar behaviour for the high-quality sample described in Sect.~\ref{sc:z_failures} (top right panel) and the cosmological sample described in Sect.~\ref{sc:purity_cosmo} (bottom left panel).

We also study the \HE-magnitude distribution of objects with a compatible redshift in both surveys ($|z_{\rm SPE} - z_{\rm DESI}| / (1+z_{\rm DESI}) < 0.003$, red). The comparison of the result with the histogram of all the objects with DESI counterparts provides a direct estimate of the dependence of the fraction of good redshifts with the \HE magnitude. For \HE$>$20, the shape of the two histograms is very similar, but with a different normalisation. As expected, the two histograms are very close for the high-purity and cosmological samples (top right and bottom left), since most of the objects in these samples have a correct SPE redshift. This is not the case for the full sample with \HE$<$22.5 and $0.9<z_{\rm SPE}<1.6$, for which a large fraction of redshifts are incorrect.

We can clarify the nature of the bump in the \HE-magnitude distribution of SPE objects with a DESI redshift below \HE = 20. Most of these sources are low-$z$ DESI sources ($\gtrsim1/2$ of them are below 0.1 and almost all below 0.9) with a probably incorrect \Euclid redshift measured between 0.9 and 1.8. It is thus not surprising that they are bright in the photometry since they are nearby and their low redshift implies that these sources have very few spectral features covered by \Euclid spectroscopy and thus unreliable SPE redshifts. In contrast, $\HE<19$ objects  actually at $z>0.9$ are rare. Paradoxically, the \HE-bright objects measured at $0.9<z_{\rm SPE}<1.6$ are thus less reliable than the fainter ones.

Finally, we computed the fraction of SPE redshifts agreeing with DESI as a function of the \HE magnitude (bottom right panel). Because of the phenomenon discussed above, the fraction is very low up to \HE$\sim$20. Above this limit, the fraction of correct redshifts is mostly stable, except for a slight decrease with increasing magnitude at the faint end. Our estimates of the success rate should thus not be too affected by the fact that only a small fraction of \HE-faint objects have a DESI counterpart. We can also note that the cosmological sample has a success rate above 80\% in all \HE$>$20 bins, which is very encouraging. However, we cannot exclude that the redshift of the sources with a DESI counterpart could be easier to measure, biasing our estimates of the success rate towards optimistic values. For instance, the median \ha\ flux of the cosmological sample is $3.5 \times 10^{-16}\,\esc$, while we obtain $4.3 \times 10^{-16}\,\esc$ if we restrict our computation to objects with a DESI redshift. This suggests that our DESI reference sample is biased towards brighter objects in \ha, but also that this bias is rather small.

\end{appendix}
\end{document}